%% file: main.tex
\documentclass[a4paper,11pt]{article}
\pdfoutput=1 

\usepackage{jheppub} 

\usepackage[T1]{fontenc} 

\usepackage{rotating}
\usepackage{bm}
\usepackage{titlesec}
\titleformat*{\subsubsection}{\bfseries\boldmath}

\title{\boldmath Impact of systematic and amplitude model correlations within and between systems of combined input: A case study with $\phi_2$ ($\alpha$)}

\author{J. Dalseno}
\affiliation{Instituto Galego de F\'{i}sica de Altas Enerx\'{i}ıas (IGFAE), Universidade de Santiago de Compostela, Santiago de Compostela, Spain}

\emailAdd{jeremy.peter.dalseno@cern.ch}

\usepackage{ifthen} 
\newboolean{uprightparticles}
\setboolean{uprightparticles}{false} 
\input{lhcb-symbols-def} 

\usepackage{lineno}

\abstract{The pursuit of experimental precision in the \CP-violating weak phase $\phi_2$ ($\alpha$) is not without its challenges, in part due to the need to combine multiple physical observables from various related decay channels, and therein lies a fundamental issue. Similarities in analysis procedures give rise to systematic correlations between the measured inputs constraining $\phi_2$ that must be taken into account to avoid bias. Specifically, in the case of the irreducible model uncertainty accompanying analyses involving the $\rho$ meson, it is demonstrated that ignoring correlations derived from its pole parameters, or indeed even treating correlations individually contained within each decay channel, can ultimately lead to a bias in $\phi_2$ of $\mathcal{O}(1^\circ)$. Correct treatment on the other hand, markedly reduces wandering of its central value as a function of the model uncertainty strength with the added dividend of a further improved overall uncertainty. Bias in the combination of $\Bz \to (\rho \pi)^0$ and $B \to \rho \rho$ is also seen to depend on the statistical strength of the former in relation to that of the model uncertainty in the latter. This work can inspire other studies into the points at which systematic correlations beyond those determined in single measurements matter in combinations leading to other \CP-violating weak phases such as $\phi_1$ ($\beta$), $\phi_3$ ($\gamma$) and $\phi_s$.}

\begin{document}
\maketitle
\flushbottom

\input{intro}
\input{isospin}
\input{nbb}
\input{rhorho}
\input{rhopi}
\input{conclusion}

\appendix
\input{covariance}

\acknowledgments

I am indebted to my colleagues, B.~Adeva Andany, V.~Chobanova, D. Martinez Santos and P.~Naik, for their assistance in providing a review for this paper. As always, I am greatly appreciative of the support from T.~Gershon, whose exchanges and careful reading also improved this work immensely. This work is supported by the ``Mar\'{i}a de Maeztu'' Units of Excellence program MDM2016-0692 and the Spanish Research State Agency. Financial support from the Xunta de Galicia (Centro singular de investigaci\'{o}n de Galicia accreditation 2019-2022) and the European Union (European Regional Development Fund – ERDF), is also gratefully acknowledged. 

\bibliographystyle{JHEP}
\bibliography{main}

\end{document}

%% file: lhcb-symbols-def.tex
\usepackage{xspace} 
\usepackage{upgreek}

\def\lhcb   {\mbox{LHCb}\xspace}

\def\belle  {\mbox{Belle}\xspace}
\def\belletwo {\mbox{Belle~II}\xspace}

\def\MagUp {\mbox{\em Mag\kern -0.05em Up}\xspace}

\ifthenelse{\boolean{uprightparticles}}%
{

 \def\Ppi         {\ensuremath{\uppi}\xspace}                 
                  
 \def\Prho        {\ensuremath{\uprho}\xspace}

 \def\PDelta      {\ensuremath{\Delta}\xspace}                 
 \def\PXi         {\ensuremath{\Xi}\xspace}                 
 \def\PLambda     {\ensuremath{\Lambda}\xspace}                 
 \def\PSigma      {\ensuremath{\Sigma}\xspace}                 
 \def\POmega      {\ensuremath{\Omega}\xspace}                 
 \def\PUpsilon    {\ensuremath{\Upsilon}\xspace}

 \def\PB      {\ensuremath{\mathrm{B}}\xspace}                 
                  
 \def\PD      {\ensuremath{\mathrm{D}}\xspace}

 \def\PK      {\ensuremath{\mathrm{K}}\xspace}

 \def\Pb      {\ensuremath{\mathrm{b}}\xspace}                 
                  
 \def\Pd      {\ensuremath{\mathrm{d}}\xspace}                 
 \def\Pe      {\ensuremath{\mathrm{e}}\xspace}

 \def\Pi      {\ensuremath{\mathrm{i}}\xspace}

 \def\Ps      {\ensuremath{\mathrm{s}}\xspace}                 
 \def\Pt      {\ensuremath{\mathrm{t}}\xspace}                 
 \def\Pu      {\ensuremath{\mathrm{u}}\xspace}

 \def\thebaroffset{0.0em}
}
{

 \def\Ppi         {\ensuremath{\pi}\xspace}                 
                  
 \def\Prho        {\ensuremath{\rho}\xspace}

 \mathchardef\PDelta="7101
 \mathchardef\PXi="7104
 \mathchardef\PLambda="7103
 \mathchardef\PSigma="7106
 \mathchardef\POmega="710A
 \mathchardef\PUpsilon="7107
                  
 \def\PB      {\ensuremath{B}\xspace}                 
                  
 \def\PD      {\ensuremath{D}\xspace}

 \def\PK      {\ensuremath{K}\xspace}

 \def\Pb      {\ensuremath{b}\xspace}                 
                  
 \def\Pd      {\ensuremath{d}\xspace}                 
 \def\Pe      {\ensuremath{e}\xspace}

 \def\Pi      {\ensuremath{i}\xspace}

 \def\Ps      {\ensuremath{s}\xspace}                 
 \def\Pt      {\ensuremath{t}\xspace}                 
 \def\Pu      {\ensuremath{u}\xspace}

 \def\thebaroffset{0.18em}
}
\newcommand{\offsetoverline}[2][\thebaroffset]{\kern #1\overline{\kern -#1 #2}}%

\makeatletter
\ifcase \@ptsize \relax
  \newcommand{\miniscule}{\@setfontsize\miniscule{4}{5}}
\or
  \newcommand{\miniscule}{\@setfontsize\miniscule{5}{6}}
\or
  \newcommand{\miniscule}{\@setfontsize\miniscule{5}{6}}
\fi
\makeatother

\DeclareRobustCommand{\optbar}[1]{\shortstack{{\miniscule (\rule[.5ex]{1.25em}{.18mm})}
  \\ [-.7ex] $#1$}}

\def\epem       {{\ensuremath{\Pe^+\Pe^-}}\xspace}

\def\uquark    {{\ensuremath{\Pu}}\xspace}
\def\uquarkbar {{\ensuremath{\overline \uquark}}\xspace}

\def\dquark    {{\ensuremath{\Pd}}\xspace}
\def\dquarkbar {{\ensuremath{\overline \dquark}}\xspace}

\def\squark    {{\ensuremath{\Ps}}\xspace}

\def\bquark    {{\ensuremath{\Pb}}\xspace}
\def\bquarkbar {{\ensuremath{\overline \bquark}}\xspace}

\def\tquark    {{\ensuremath{\Pt}}\xspace}

\def\pion   {{\ensuremath{\Ppi}}\xspace}
\def\piz    {{\ensuremath{\pion^0}}\xspace}
\def\pip    {{\ensuremath{\pion^+}}\xspace}
\def\pim    {{\ensuremath{\pion^-}}\xspace}
\def\pipm   {{\ensuremath{\pion^\pm}}\xspace}
\def\pimp   {{\ensuremath{\pion^\mp}}\xspace}

\def\rhomeson {{\ensuremath{\Prho}}\xspace}
\def\rhoz     {{\ensuremath{\rhomeson^0}}\xspace}
\def\rhop     {{\ensuremath{\rhomeson^+}}\xspace}
\def\rhom     {{\ensuremath{\rhomeson^-}}\xspace}

\def\KorKbar {\kern \thebaroffset\optbar{\kern -\thebaroffset \PK}{}\xspace}

\def\D       {{\ensuremath{\PD}}\xspace}

\def\DorDbar {\kern \thebaroffset\optbar{\kern -\thebaroffset \PD}\xspace}

\def\Dp      {{\ensuremath{\D^+}}\xspace}
\def\Dm      {{\ensuremath{\D^-}}\xspace}

\def\DpDm    {\ensuremath{\Dp {\kern -0.16em \Dm}}\xspace}

\def\B       {{\ensuremath{\PB}}\xspace}
\def\Bbar    {{\ensuremath{\offsetoverline{\PB}}}\xspace}

\def\BorBbar {\kern \thebaroffset\optbar{\kern -\thebaroffset \PB}\xspace}
\def\Bz      {{\ensuremath{\B^0}}\xspace}
\def\Bzb     {{\ensuremath{\Bbar{}^0}}\xspace}
\def\Bd      {{\ensuremath{\B^0}}\xspace}

\def\BdorBdbar {\kern \thebaroffset\optbar{\kern -\thebaroffset \Bd}\xspace}
\def\Bu      {{\ensuremath{\B^+}}\xspace}
\def\Bub     {{\ensuremath{\B^-}}\xspace}
\def\Bp      {{\ensuremath{\Bu}}\xspace}
\def\Bm      {{\ensuremath{\Bub}}\xspace}

\def\Bs      {{\ensuremath{\B^0_\squark}}\xspace}

\def\BsorBsbar {\kern \thebaroffset\optbar{\kern -\thebaroffset \Bs}\xspace}

\def\Y#1S{\ensuremath{\PUpsilon{(#1S)}}\xspace}

\def\LorLbar     {\kern \thebaroffset\optbar{\kern -\thebaroffset \PLambda}\xspace}

\def\BF         {{\ensuremath{\mathcal{B}}}\xspace}

\def\to                 {\ensuremath{\rightarrow}\xspace}

\def\CP                {{\ensuremath{C\!P}}\xspace}

\def\Vud  {{\ensuremath{V_{\uquark\dquark}^{\phantom{\ast}}}}\xspace}

\def\Vtd  {{\ensuremath{V_{\tquark\dquark}^{\phantom{\ast}}}}\xspace}

\def\Vubs  {{\ensuremath{V_{\uquark\bquark}^\ast}}\xspace}

\def\Vtbs  {{\ensuremath{V_{\tquark\bquark}^\ast}}\xspace}

\def\AT#1     {\ensuremath{A_{\mathrm{T}}^{#1}}\xspace}           

\def\C#1      {\ensuremath{\mathcal{C}_{#1}}\xspace}                       
\def\Cp#1     {\ensuremath{\mathcal{C}_{#1}^{'}}\xspace}                    
\def\Ceff#1   {\ensuremath{\mathcal{C}_{#1}^{\mathrm{(eff)}}}\xspace}        
\def\Cpeff#1  {\ensuremath{\mathcal{C}_{#1}^{'\mathrm{(eff)}}}\xspace}       
\def\Ope#1    {\ensuremath{\mathcal{O}_{#1}}\xspace}                       
\def\Opep#1   {\ensuremath{\mathcal{O}_{#1}^{'}}\xspace}                    

\newcommand{\aunit}[1]{\ensuremath{\text{\,#1}}}       

\newcommand{\tev}{\aunit{Te\kern -0.1em V}\xspace}
\newcommand{\gev}{\aunit{Ge\kern -0.1em V}\xspace}
\newcommand{\mev}{\aunit{Me\kern -0.1em V}\xspace}
\newcommand{\kev}{\aunit{ke\kern -0.1em V}\xspace}
\newcommand{\ev}{\aunit{e\kern -0.1em V}\xspace}
 
\newcommand{\mevc}{\ensuremath{\aunit{Me\kern -0.1em V\!/}c}\xspace}
\newcommand{\gevc}{\ensuremath{\aunit{Ge\kern -0.1em V\!/}c}\xspace}
\newcommand{\mevcc}{\ensuremath{\aunit{Me\kern -0.1em V\!/}c^2}\xspace}
\newcommand{\gevcc}{\ensuremath{\aunit{Ge\kern -0.1em V\!/}c^2}\xspace}

\def\ab   {\ensuremath{\aunit{ab}}\xspace}

\def\gsim{{~\raise.15em\hbox{$>$}\kern-.85em
          \lower.35em\hbox{$\sim$}~}\xspace}
\def\lsim{{~\raise.15em\hbox{$<$}\kern-.85em
          \lower.35em\hbox{$\sim$}~}\xspace}

\def\tell1  {TELL1\xspace}
\def\ukl1   {UKL1\xspace}



%% file: intro.tex
\section{\boldmath Introduction}
\label{sec:intro}

Violation of the combined charge-parity symmetry~(\CP violation) in the Standard Model~(SM) arises from a single irreducible phase in the Cabibbo-Kobayashi-Maskawa~(CKM) quark-mixing matrix~\cite{Cabibbo:1963yz,Kobayashi:1973fv}. Various processes offer different yet complementary insight into this phase, which manifests in a number of experimental observables over-constraining the Unitarity Triangle (UT). The measurement of such parameters and their subsequent combination is important as New Physics~(NP) contributions can present themselves as an inconsistency within the triangle paradigm.
\begin{figure}[!htb]
    \centering
    \includegraphics[height=115pt,width=!]{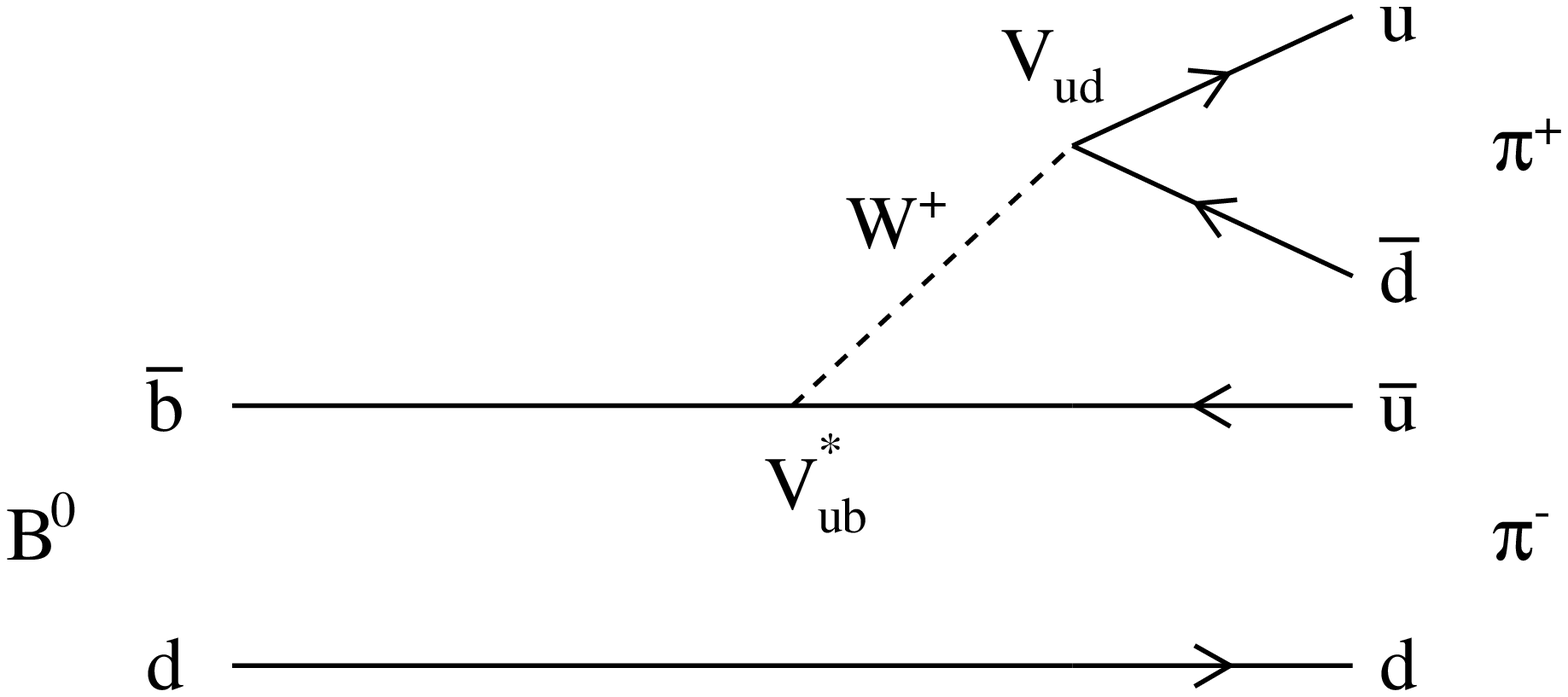}
    \includegraphics[height=115pt,width=!]{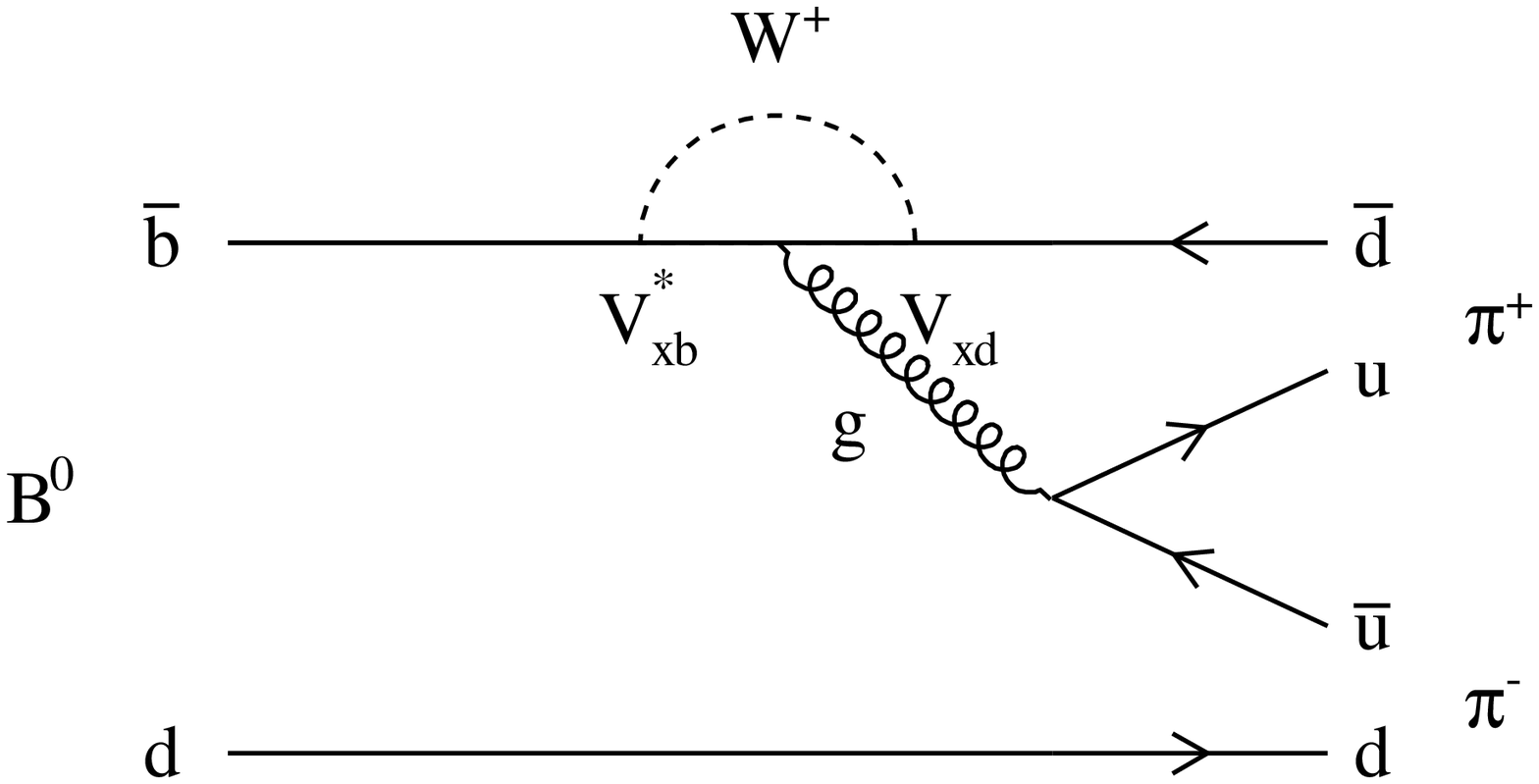}
    \put(-425,105){(a)}
    \put(-210,105){(b)}
    \caption{\label{fig:pipi} Leading-order Feynman diagrams shown producing $\Bz \to \pip \pim$ decays, though the same quark transition can also produce $\Bz \to \rho^{\pm} \pimp$, $\rho^+ \rho^-$ and $a_1^\pm \pimp$. (a) depicts the dominant (tree) diagram while (b) shows the competing loop (penguin) diagram. In the penguin diagram, the subscript $x$ in $V_{xb}$ refers to the flavour of the intermediate-state quark $(x=u,c,t)$.}
\end{figure}

Decays that proceed predominantly through the $\bquarkbar \to \uquarkbar \uquark \dquarkbar$ tree transition (figure~\ref{fig:pipi}a) in the presence of \Bz--\Bzb mixing are sensitive to the interior angle of the UT, $\phi_2 = \alpha \equiv \arg(-\Vtd \Vtbs)/(\Vud \Vubs)$, which can be accessed through mixing-induced \CP violation observables measured from time-dependent, flavour-tagged analyses. This quark
process manifests itself in multiple systems, including $B \to \pi \pi$~\cite{Lees:2012mma,Adachi:2013mae,Aaij:2018tfw,Aaij:2020buf,Aubert:2007hh,Duh:2012ie,Julius:2017jso}, $(\rho \pi)^0$~\cite{Lees:2013nwa,Kusaka:2007dv,Kusaka:2007mj}, $\rho \rho$~\cite{Aubert:2007nua,Vanhoefer:2015ijw,Aubert:2009it,Zhang:2003up,Aubert:2008au,Adachi:2012cz,Aaij:2015ria} and $a_1^\pm \pi^\mp$~\cite{Aubert:2006gb,Dalseno:2012hp,Aubert:2009ab}, where the angle $\phi_2$ has so far been constrained with an overall uncertainty of around $4^\circ$~\cite{Gronau:2016idx,Charles:2017evz,Bona:2006ah,Amhis:2019ckw}. With the dubious honour of being the least known input to the UT now falling to $\phi_2$, there has never been better motivation to improve its experimental precision.

More often than not, this involves the combination of several physics parameters extracted from related decay channels raising an important general question that hitherto has not yet been explored in any great detail. As experimental measurements become more and more precise, a crucial unknown is the point at which it will become necessary to consider systematic correlations, not arising simply within individual analyses, but rather in between the relevant analyses in order to avoid bias. As a specific case study of a broader issue, which includes but is not limited to combination-based approaches designed to measure other \CP-violating weak phases such as $\phi_1$ ($\beta$)~\cite{Ciuchini:2005mg,Faller:2008zc,Ciuchini:2011kd,Jung:2012mp,DeBruyn:2014oga,Frings:2015eva,Ligeti:2015yma,Barel:2020jvf}, $\phi_3$ ($\gamma$)~\cite{Lorier:2010xf,Imbeault:2010xg,Rey-LeLorier:2011ltd} and $\phi_s$~\cite{Faller:2008zc,DeBruyn:2014oga,Barel:2020jvf,Fleischer:1999zi,Faller:2008gt,Liu:2013nea}, I open the discussion in this paper within the context of the $\phi_2$ average. By and large, this problem is generally an internal matter for each collaboration, however there are irreducible systematic uncertainties that transcend experiment, warranting a more cooperative approach and thus is the primary focus here. Experience in amplitude analysis suggests that the model uncertainty of a dominant vector resonance tends to derive more significantly from its own pole parameters rather than the remainder of the model, which is converse to smaller contributions where the opposite trend appears to hold. This is because Breit-Wigner phases vary most rapidly at the poles, exacerbating the effect of small variations to manipulate interference patterns in the regions of greatest physical interest. In this specific consideration, these pole parameters are those of the $\rho$ meson.

I open in section~\ref{sec:isospin}, with a description of the SU(2)-based approach for controlling distortions in experimental $\phi_2$ measurements arising from the ever-present strong-loop gluonic penguin processes. Following this, I introduce the impact of systematic correlations in section~\ref{sec:nbb} with a conceptually simpler example surrounding the branching fractions of the decay processes involved. In section~\ref{sec:rhorho}, I move into the primary study on the bias in $\phi_2$ caused by neglecting amplitude model correlations in the $B \to \rho \rho$ system arising from the $\rho$ pole masses and widths. This bias, if left unchecked, can then go on to affect the otherwise immune $B^0 \to (\rho \pi)^0$ analysis as explained in section~\ref{sec:rhopi}. Finally, conclusions are drawn in section~\ref{sec:conclusion} along with some recommendations on how the community going forward can reduce $\phi_2$ bias induced by systematic and amplitude model correlations.

%% file: isospin.tex
\section{\boldmath Strong-penguin containment in $\phi_2$ constraints}
\label{sec:isospin}

In general, the extraction of $\phi_2$ is complicated by the presence of interfering amplitudes that distort the experimentally determined value of $\phi_2$ from its SM expectation and would mask any NP phase if not accounted for. These effects primarily include $\bquarkbar \rightarrow \dquarkbar u \uquarkbar$ strong-loop decays (figure~\ref{fig:pipi}b), although isospin-violating processes such as electroweak penguins, $\piz$--$\eta$--$\eta^\prime$ mixing, $\rho^0$--$\omega$--$\phi$ mixing~\cite{Gronau:2005pq} and the finite $\rho$ width in $B \to \rho\rho$~\cite{Falk:2003uq} can also play a role.

\subsection{Original approach}

It is possible to remove the isospin-conserving component of this contamination by invoking SU(2) arguments. The original method considers the three possible charge configurations of $B \rightarrow \pi\pi$ decays~\cite{Gronau:1990ka}. Bose-Einstein statistics rules out a total isospin $I=1$ contribution, leaving just the $I=0, 2$ amplitudes remaining. Strong penguins then only have the possibility to contribute an $I=0$ amplitude, since the mediating gluon is an isospin singlet. However, in the specific case of $\Bp \rightarrow \pip \piz$, the further limiting projection $I_{3} = 1$ additionally rules out $I=0$, thereby forbidding strong penguin contributions to this channel.
\begin{figure}[!htb]
    \centering
    \includegraphics[height=110pt,width=!]{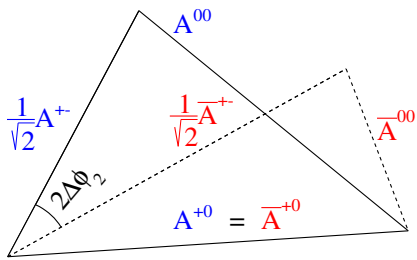}
    \caption{\label{fig_iso} Complex isospin amplitude triangles from which $\Delta \phi_2$ can be determined.}
\end{figure}

The complex $B \rightarrow \pi\pi$ and $\bar B \rightarrow \pi\pi$ decay amplitudes obey the isospin relations
\begin{equation}
    A^{+0} = \frac{1}{\sqrt{2}}A^{+-} + A^{00}, \;\;\;\; \bar{A}^{+0} = \frac{1}{\sqrt{2}}\bar{A}^{+-} + \bar{A}^{00},
  \label{eq_iso}
\end{equation}
respectively, where the superscripts refer to the combination of pion charges. The decay amplitudes can be represented as triangles in the complex plane as shown in figure~\ref{fig_iso}. As $\Bp \rightarrow \pip \piz$ is a pure tree mode, its amplitude in isospin space is identical to that of its \CP-conjugate and so these triangles lose their relative orientation to share the same base, $A^{+0}=\bar{A}^{+0}$, allowing the shift in $\phi_2$ caused by strong penguin contributions $\Delta \phi_2 \equiv \phi_2^\pm - \phi_2$, to be determined from the phase difference between $\bar{A}^{+-}$ and $A^{+-}$. These amplitudes can be constrained by 7 mostly independent physical observables for a two-fold discrete ambiguity in the range $[0, 180]^\circ$, which are related to the decay amplitudes as
\begin{equation}
    \label{eq:old}
    \frac{1}{\tau_B^{i+j}} \BF^{ij} = \frac{|\bar A^{ij}|^2 + |A^{ij}|^2}{2}, \hspace{10pt} \mathcal{A}_{\CP}^{ij} = \frac{|\bar A^{ij}|^2 - |A^{ij}|^2}{|\bar A^{ij}|^2 + |A^{ij}|^2}, \hspace{10pt} \mathcal{S}_{\CP}^{ij} = \frac{2\Im(\bar A^{ij} A^{ij*})}{|\bar A^{ij}|^2 + |A^{ij}|^2},
\end{equation}
where \BF, $\mathcal{A}_{\CP}$ and $\mathcal{S}_{\CP}$ are the branching fractions, \CP violation in the decay and mixing-induced \CP violation parameters, respectively. The superscript $ij$, represents the charge configuration of the final state pions and $\tau_B^{i+j}$ is the lifetime of the \Bp~($i+j=1$) or \Bz~($i+j=0$). Naturally for $\Bp \rightarrow \pip \piz$, \CP violation in the decay is forbidden by the isospin argument and mixing-induced \CP violation is not defined. The ambiguity in $\phi_2$ is also increased to 8-fold if $\mathcal{S}_{\CP}^{00}$ of the colour-suppressed channel is not measured as is currently the case. This approach is also applied to the $B \to \rho \rho$ system analogously, substituting the $\rho$ meson in place of each pion.

\subsection{Next-generation approach}

The $B \to \rho \rho$ system presents a greater theoretical and experimental challenge over $B \to \pi \pi$. It has already been pointed out that isospin-breaking ($I=1$) $\rho$-width effects can be controlled by reducing the $\rho$ analysis window of $\Bz \to \rho^+ \rho^-$ and $\Bp \to \rho^+ \rho^0$ according to the method outlined in ref.~\cite{Gronau:2016nnc}. An open question to be studied is the extent to which this is systematically feasible in the presence of interfering and non-interfering backgrounds.

In this work, I espouse an alternate viewpoint in which the possibility to exploit the multi-body final state through directly modelling the structure of $\rho^0$--$\omega$ mixing and $I=1$ finite $\rho$-width effects is acquired in exchange for greater analysis complexity. To that end, I have already outlined the amplitude analysis framework by which this can be achieved, replacing the measured physical observables from eq.~\ref{eq:old} by
\begin{equation}
    \label{eq:new}
    \frac{1}{\tau_B^{i+j}} \BF^{ij} = \frac{|\bar A^{ij}|^2 + |A^{ij}|^2}{2}, \hspace{10pt} |\lambda_{\CP}^{ij}| = \biggl|\frac{\bar A^{ij}}{A^{ij}}\biggr|, \hspace{10pt} \phi_2^{ij} = \frac{\arg(\bar A^{ij} A^{ij*})}{2},
\end{equation}
where $\lambda^{ij}_{\CP}$ is a \CP-violation parameter and $\phi_2^{ij}$ is its effective weak phase. As these quantities are now related to the isospin triangles at amplitude level, the solution degeneracy in $\phi_2$ for the range $[0, 180]^\circ$ is resolved~\cite{Dalseno:2018hvf} and as an added incentive, the 8-fold solution degeneracy in $B^0 \to a_1^\pm \pimp$ can also be lifted for the same range in the SU(3) approach~\cite{Dalseno:2019kps}. Naturally, this method raises concerns regarding the potential impact on $\phi_2$ coming from correlated amplitude model systematics which will be studied here.

\subsection{Statistical method}

In this paper, I employ the frequentist approach adopted by the CKMfitter Group~\cite{Charles:2017evz} where a $\chi^2$ is constructed comparing theoretical forms for physical observables expressed in terms of parameters of interest, $\bm{\mu}$, with their experimentally measured values, $\bm{ x}$. The most general form,
\begin{equation}
    \chi^2 \equiv (\bm{x}-\bm{\mu})^T \bm{\Sigma}^{-1} (\bm{x}-\bm{\mu}),
\end{equation}
is necessary here, where $\bm{\Sigma}$ is total covariance matrix composed of the statistical and systematic covariance matrices as $\bm{\Sigma} \equiv \bm{\Sigma}_{\rm Stat} + \bm{\Sigma}_{\rm Syst}$. The statistical covariance matrix comes directly from the function minimisation procedure during the nominal fit to a sample, while the systematic covariance matrix is manually derived. Parameter variations are generated according to their uncertainties and the fit is repeated for each set of variations. Over $N$ fits, the covariance between a pair of physical observables is given by
\begin{equation}
    \Sigma_{x,y} \equiv \sum^N_{i=1} \frac{(x_i-\bar x)(y_i-\bar y)}{N},
\end{equation}
where the barred quantities representing the means are obtained from the nominal fit.

A scan is then performed, minimising the $\chi^2$ to determine $\bm{\mu}$ for each value of $\phi_2$ fixed across a range. The value of $\Delta \chi^2$ from the global minimum is finally converted into a $p$-value scan, assuming it is distributed with one degree of freedom, from which confidence intervals can be derived.

%% file: nbb.tex
\section{\boldmath Systematic correlations within systems}
\label{sec:nbb}

Before delving into the main point regarding amplitude model correlations, it may be advantageous to introduce this difficult topic by digressing to conceptually simpler systematic correlations that can be trivially accounted for here. One such example is the number of $B \Bbar$ pairs produced at $\epem$ machines operating at the $\Upsilon(4S)$ resonance, $N_{B\Bbar}$, that enters the absolute branching fraction calculations in $B \to \pi \pi$ decays through
\begin{equation}
    \BF^{ij} = \frac{N^{ij}}{\epsilon^{ij}N_{B\Bbar}},
\end{equation}
where $N^{ij}$ is the extracted signal yield and $\epsilon^{ij}$ is the reconstruction efficiency of that mode. 
Although equal production of \Bp \Bm and \Bz \Bzb pairs is implicitly assumed here for simplicity, this will need to be evaluated at \belletwo as the current uncertainties on their rates~\cite{ParticleDataGroup:2020ssz} would otherwise constitute the dominant systematic instead of those arising from $N_{B\Bbar}$.
It can immediately be seen that all three branching fractions are 100\% systematically correlated in $N_{B\Bbar}$, because as a quantity that is independent of the channel being studied, whatever direction it fluctuates in, all branching fractions must follow suit by the same factor.

\begin{table}[!htb]
    \centering
    \begin{tabular}{|c|c|}
    \hline
    Parameter & \belletwo projection\\ \hline
    $\BF(\pip \piz)$ ($10^{-6}$) & $\phantom{+}5.86 \pm 0.03 \pm 0.09$\\
    $\BF(\pip \pim)$ ($10^{-6}$) & $\phantom{+}5.04 \pm 0.03 \pm 0.08$\\
    $\BF(\piz \piz)$ ($10^{-6}$) & $\phantom{+}1.31 \pm 0.03 \pm 0.03$\\
    $\mathcal{A}_{\CP}(\pip \pim)$ & $+0.33 \pm 0.01 \pm 0.03$\\
    $\mathcal{S}_{\CP}(\pip \pim)$ & $-0.64 \pm 0.01 \pm 0.01$\\
    $\mathcal{A}_{\CP}(\piz \piz)$ & $+0.14 \pm 0.03 \pm 0.01$\\ \hline
    \end{tabular}
    \caption{Projections for $B \to \pi \pi$ physics observables with $50 \ab^{-1}$ taken from ref.~\cite{Kou:2018nap} where the first uncertainty is statistical and the second is systematic.}
  \label{tab:nbb}
\end{table}
To illustrate, I repeat the $\phi_2$ projection for \belletwo with $50 \ab^{-1}$ with and without accounting for systematic correlation arising from $N_{B\Bbar}$. Input is borrowed from ref.~\cite{Kou:2018nap} and displayed verbatim in table~\ref{tab:nbb}. As the systematic uncertainty is considered to be irreducible and kept at the $1.37\%$ level from \belle, it is the dominant expected systematic by far. For the purposes of demonstrating impact on $\phi_2$, I will then assume that the branching fraction systematics are entirely due to the uncertainty in $N_{B\Bbar}$, and thus the systematic correlation matrix can be immediately written down as shown in table~\ref{tab:nbbcorr}. The only known statistical correlation is between $\mathcal{A}_{\CP}(\pip \pim)$ and $\mathcal{S}_{\CP}(\pip \pim)$, set at $+0.10$ from the Belle result.
\begin{table}[!htb]
    \centering
    \begin{tabular}{|c|cccccc|}
    \hline
    & $\BF(\pip \piz)$ & $\BF(\pip \pim)$ & $\BF(\piz \piz)$ & $\mathcal{A}_{\CP}(\pip \pim)$ & $\mathcal{S}_{\CP}(\pip \pim)$ & $\mathcal{A}_{\CP}(\piz \piz)$\\ \hline
    $\BF(\pip \piz)$ & $+1$ & & & & &\\
    $\BF(\pip \pim)$ & $+1$ & $+1$ & & & &\\
    $\BF(\piz \piz)$ & $+1$ & $+1$ & $+1$ & & &\\
    $\mathcal{A}_{\CP}(\pip \pim)$ & $\phantom{+}0$ & $\phantom{+}0$ & $\phantom{+}0$ & $+1$ & &\\
    $\mathcal{S}_{\CP}(\pip \pim)$ & $\phantom{+}0$ & $\phantom{+}0$ & $\phantom{+}0$ & $\phantom{+}0$ & $+1$ &\\
    $\mathcal{A}_{\CP}(\piz \piz)$ & $\phantom{+}0$ & $\phantom{+}0$ & $\phantom{+}0$ & $\phantom{+}0$ & $\phantom{+}0$ & $+1$\\ \hline
    \end{tabular}
    \caption{Systematic correlation matrix between $B \to \pi \pi$ physics observables assuming only the uncertainty in $N_{B\Bbar}$ contributes.}
  \label{tab:nbbcorr}
\end{table}

\begin{figure}[!htb]
    \centering
    \includegraphics[height=150pt,width=!]{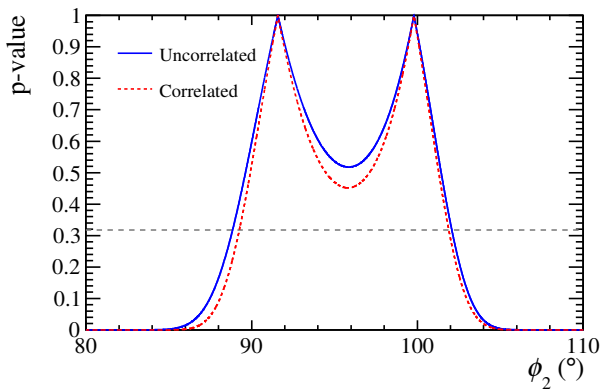}
    \caption{$p$-value scan of $\phi_2$ where the horizontal dashed line shows the $1\sigma$ bound. The blue curve shows the constraint ignoring systematic correlations, while the red considers them in $N_{B\Bbar}$.}
    \label{fig:nbb}
\end{figure}

The $\phi_2$ scan is then conducted in the vicinity of the SM solution with and without systematic correlations, the results of which can be seen in figure~\ref{fig:nbb}. The leading edge of the solution consistent with the SM is seen to improve by $0.4^\circ$ when accounting for systematic correlations, a striking result within the context of the sub-degree precision anticipated at \belletwo. At a first glance, this may seem counter-intuitive as some may recall the familiar summation of correlated uncertainties linearly over the more favourable summation in quadrature for uncorrelated cases. However, this is more applicable to instances of single physics parameters, whereas between physics parameters, correlations restrict statistical freedom. In this example, as the uncertainty in $N_{B\Bbar}$ is not allowed to nonsensically follow three independent statistical distributions, as would be encapsulated by the identity correlation matrix, the $\phi_2$ constraint improves in consequence.

Although the knowledge of $N_{B\Bbar}$ is expected to dominate the systematic uncertainty on the branching fractions, ideally the full systematic covariance matrix will be constructed in future analyses considering all sources. For example, as common control samples provide the tracking, particle identification and \piz\ reconstruction uncertainties, the branching fractions are again systematically correlated in these categories. Concurrently, the \CP-violating parameters are also affected with the timing resolution and flavour-tagging performance being obtained from common studies. However, perhaps the most dangerous systematic here would be the shared method for evaluating tag-side interference from doubly-Cabibbo-suppressed decays~\cite{Long:2003wq}, which is also considered to be another irreducible systematic up until the point where it becomes statistically advantageous to rely exclusively on semileptonic flavour tags.




%% file: rhorho.tex
\section{\boldmath Amplitude model correlations within systems}
\label{sec:rhorho}

Although amplitude analysis has seen limited involvement~\cite{Aaij:2015ria} in the $B \to \rho \rho$ constraint of $\phi_2$, it stands to reason that this approach will become more attractive in controlling uncertainties as data samples increase. For the small cost of modelling one additional variable over current analyses, the necessary degrees of freedom are acquired to harness the full statistics particularly of the limiting colour-suppressed $\Bz \to \rhoz \rhoz$ decay, thereby improving $\phi_2$ precision in this sector, and even opening the possibility to determine $\phi_2$ separately for each of the three polarisation configurations of $B \to \rho \rho$. Furthermore, amplitude analysis allows the direct modelling of interfering components such as the dipion $I=0, 1$ resonant contributions and other non-resonant S-wave effects such as elastic and inelastic particle rescattering processes in the vicinity of the $\rho$, thus reducing model uncertainty estimations. Perhaps most importantly, isospin-breaking contributions known to bias $\phi_2$ can also be accounted for in the amplitude model, such as with the $\rho^0$--$\omega$ mixing lineshape of refs.~\cite{Aaij:2019hzr,Aaij:2019jaq} and the structure of $I=1$ finite $\rho$-width effects suggested in ref.~\cite{Falk:2003uq}.

One aspect these three analyses have in common is fixed $\rho$ pole parameters, so therein lies the potential for systematic model uncertainties to impact the $\phi_2$ average. Unlike the $N_{B\Bbar}$ case mentioned in the previous section, these are not multiplicative factors to any physics parameter and as such the correlation matrix cannot immediately be written down. In order for the covariance matrix to be derived, repeated randomised systematic variations on a sample of each $B \to \rho \rho$ channel needs to be applied. Then to generate these samples, amplitude models are first required for which information is sparse, meaning that assumptions will have to be made on the magnitudes and relative phases between the $B \to \rho \rho$ polarisations. In previous works~\cite{Dalseno:2018hvf,Dalseno:2019kps}, unknown physics parameters were uniformly distributed in an ensemble test to give a sense of what to expect on average in their respective $\phi_2$ studies. However in this case, applying systematic variations on top of all three amplitude analyses in an ensemble is not a practical endeavour and would be of unclear benefit, besides. Therefore, conclusions from this paper will be limited to identifying the scale of potential bias in $\phi_2$ induced by neglecting amplitude model correlations, as opposed to providing a definitive range.

\subsection{Amplitude model}

Yields are set based on \belle results according to expectations for $50 \ab^{-1}$ to be collected with \belletwo. I consider rudimentary models with contributions to the 4-body phase space coming only from the channels known to exist in the analysis region. The amplitude for each intermediate state at position $\Phi_4$, is parameterised as
\begin{equation}
    A_i(\Phi_4) = B^L_B(\Phi_4) \cdot [B^L_{R_1}(\Phi_4) T_{R_1} (\Phi_4)] \cdot [B^L_{R_2}(\Phi_4) T_{R_2} (\Phi_4)] \cdot S_i(\Phi_4),
\end{equation}
where $B^L_B$ represents the production Blatt-Weisskopf barrier factor~\cite{VonHippel:1972fg} depending on the orbital angular momentum between the products of the $B$ decays, $L$. Two resonances will appear in each isobar, denoted by $R_1$ and $R_2$, for which respective decay barrier factors are also assigned. The Breit-Wigner propagators are represented by $T$, while the overall spin amplitude is given by $S$. Each isobar is Bose-symmetrised as necessary so that the total amplitude is always symmetric under the exchange of like-sign pions.

The Blatt-Weisskopf penetration factors account for the finite size of the decaying resonances by assuming a square-well interaction potential with radius $r$. They depend on the breakup momentum between the decay products $q$, and the orbital angular momentum between them $L$. Their explicit expressions used in this analysis are
\begin{eqnarray}
    B^0(q) &=& 1, \nonumber \\
    B^1(q) &=& \frac{1}{\sqrt{1+(qr)^2}}, \nonumber\\
    B^2(q) &=& \frac{1}{\sqrt{9+3(qr)^2+(qr)^4}}.
\end{eqnarray}

Spin amplitudes are constructed with the covariant tensor formalism based on the Rarita-Schwinger conditions~\cite{Rarita:1941mf}. The spin $S$, of some state with 4-momentum $p$, and spin projection $s_z$, is represented by a rank-$S$ polarisation tensor that is symmetric, traceless and orthogonal to $p$. These conditions reduce the number of independent elements to $2S+1$ in accordance with the number of degrees of freedom available to a spin-$S$ state. The sum over these polarisation indices of the inner product of polarisation tensors form the fundamental basis on which all spin amplitudes are built. Called the spin projection operator $P$, it projects an arbitrary tensor onto the subspace spanned by the spin projections of the spin-$S$ state.

Another particularly useful object is the relative orbital angular momentum spin tensor $L$, which for some process $R \to P_1 P_2$, is the relative momenta of the decay products $q_R \equiv p_1 - p_2$ projected to align with the spin of $R$,
\begin{equation}
    \label{eq:orbital}
    L_{\mu_1 \mu_2 ... \mu_L}(p_R, q_R) = P_{\mu_1 \mu_2 ... \mu_L \nu_1 \nu_2 ... \nu_L} (p_R) q_R^{\nu_1} q_R^{\nu_2} ... q_R^{\nu_L},
\end{equation}
where the number of indices representing the tensor rank is equal to the value of $L$. Finally, to ensure that the spin amplitude behaves correctly under parity transformation, it is sometimes necessary to include the Levi-Civita totally antisymmetric tensor $\epsilon_{abcd}p_R^d$. Each stage of a decay is represented by a Lorentz scalar obtained by contracting an orbital tensor between the decay products with a spin wavefunction of equal rank representing the final state. Three spin topologies are necessary for $B \to \rho \rho$ as $S$-, $P$- and $D$-waves are permitted between the vector resonances, with total spin densities,
\begin{eqnarray}
    \label{eq:spin}
    &&S\text{-wave}: \hspace{10pt} S \propto L_a(p_{\rho_1}, q_{\rho_1})L^a(p_{\rho_2}, q_{\rho_2}), \nonumber\\
    &&P\text{-wave}: \hspace{10pt} S \propto \epsilon_{abcd} L^d(p_{B}, q_{B}) L^c(p_{\rho_1}, q_{\rho_1}) L^b(p_{\rho_2}, q_{\rho_2}) p^a_{B}, \nonumber\\
    &&D\text{-wave}: \hspace{10pt} S \propto L_{ab}(p_{B}, q_{B}) L^b(p_{\rho_1}, q_{\rho_1}) L^a(p_{\rho_2}, q_{\rho_2}).
\end{eqnarray}

In general, resonance lineshapes are described by Breit-Wigner propagators as a function of the energy-squared $s$,
\begin{equation}
    T(s) = \frac{1}{M^2(s) - s - i\sqrt{s}\Gamma(s)},
\end{equation}
where $M^2(s)$ is the energy-dependent mass and $\Gamma(s)$ is the total width which is normalised such that it represents the nominal width $\Gamma_0$, at the pole mass, $m_0$. For the \rhoz\ resonance, the Gounaris-Sakurai parameterisation is used to provide an analytic expression for $M^2(s)$ and $\Gamma(s)$~\cite{Gounaris:1968mw}.

\subsection{Pseudo-experiment generation method}

The unknown strong complex couplings between contributions in the amplitude model are partly inspired by reverse-engineering the known branching fractions for each polarisation. The Monte Carlo (MC) is based on the decay rates in phase space, which for $\Bp \to \rhop \rhoz$ is
\begin{equation}
    \Gamma(q) = \frac{1 + q}{2} |A|^2 + \frac{1 - q}{2} |\bar A|^2,
\end{equation}
where $q = +1 (-1)$ for \Bp\ (\Bm). On the other hand, the time-dependent decay rates of \Bz\ and \Bzb\ decays to a self-conjugate final states are given by
\begin{eqnarray}
    \label{eq:tdrate}
    \Gamma(t) &\propto& e^{-t/\tau}[(|A|^2 + |\bar A|^2) + (|A|^2 - |\bar A|^2)\cos \Delta m_d t - 2 \Im (\bar A A^*) \sin \Delta m_d t], \nonumber\\
    \bar \Gamma(t) &\propto& e^{-t/\tau}[(|A|^2 + |\bar A|^2) - (|A|^2 - |\bar A|^2)\cos \Delta m_d t + 2 \Im (\bar A A^*) \sin \Delta m_d t],
\end{eqnarray}
respectively, where $A$ is the static decay amplitude, $\tau$ is the \Bz\ lifetime and $\Delta m_d$ is the mass difference between the $B_H$ and $B_L$ mass eigenstates. This form assumes no \CP violation in the mixing $|q/p| = 1$, and that the total decay rate difference between the two mass eigenstates is negligible.

The total amplitude $A$, can be written in the typical isobar approach as the coherent sum over the number of intermediate states in the model with amplitude $A_i$, as a function of 4-body phase space position $\Phi_4$,
\begin{equation}
    A \equiv \sum_i a_iA_i(\Phi_4),
\end{equation}
where $a_i$ is a strong complex coupling determined directly from the data. Incorporating a complex \CP violation parameter $\lambda_i$, for each weak contribution in the phase space, the total $\bar A$ can be written as
\begin{equation}
    \bar A \equiv \sum_i a_i \lambda_i \bar A_i(\bar \Phi_4) = \sum_i a_i \lambda_i A_i(\bar \Phi_4),
\end{equation}
for $\Bp \to \rhop \rhoz$, where the phase space of the \CP-conjugated process $\bar \Phi_4$, is set by convention to have the same sign as $\Phi_4$ for all amplitude contributions, leaving $A_i$ to contain only strong dynamics blind to flavour. Conversely, for $\Bz \to \rhop \rhom$ and $\rhoz \rhoz$,
\begin{equation}
    \bar A \equiv \sum_i a_i \lambda_i \bar A_i(\Phi_4) = \sum_i a_i \lambda_i A_i(\bar \Phi_4),
\end{equation}
the phase space of the \CP-conjugated process $\bar \Phi_4$, must be transformed relative to the elected particle ordering under $C$ and $P$ conjugation in order to achieve $A_i$ containing only strong dynamics.

Complex couplings are then evaluated through a $\chi^2$ fit relating the observed branching fractions for each isobar scaled to unity, to the fit fractions of each isobar calculated for the generated model in the 4-body phase space,
\begin{equation}
    {\cal F}^{\rm pred}_i = \frac{\int (|A_i|^2 + |\bar A_i|^2) d\Phi_4}{\int \sum_i(|A_i|^2 + |\bar A_i|^2) d\Phi_4}.
\end{equation}
The branching fractions for each polarisation are set mostly with HFLAV input~\cite{Amhis:2019ckw} except where mentioned. As the longitudinal polarisation is known to dominate, the remainder is assigned exclusively to the $P$- or \CP-odd P-wave for simplicity, while the longitudinal component is divided evenly between the $P$- or \CP-even S- and D-waves for the flavour-specific and \CP-conjugate final states, respectively. Naturally, there are 2 solutions for each free strong coupling, so whichever solution the fit converges to first is taken to generate the MC sample for each $B \to \rho \rho$ channel. Position in phase space is provided by the \texttt{GENBOD} algorithm~\cite{James:1968gu} and \texttt{qft++} gives the spin densities~\cite{Williams:2008wu}.

\subsubsection{$\Bp \to \rhop \rhoz$}

This amplitude analysis should be rather straight-forward as the phase space can be restricted to limit contributions from the $a_1(1260)$ resonances. Here, the yield is set to 100k events and the dipion range is restricted to be below the typical $1.1 \gevcc$. The input branching fractions with the determined couplings are shown in table~\ref{tab:rhoprhoz}. Note that here and throughout, the spin amplitudes given in eq.~\ref{eq:spin} are not normalised over the phase space, so there is no direct relation between the fitted couplings and their corresponding branching fractions. This also means that the relative strengths of each partial wave cannot be inferred from the couplings either as each spin factor contains different momentum scales by eq.~\ref{eq:orbital}, depending on the number of orbital angular momentum tensors involved.
\begin{table}[!htb]
    \centering
    \begin{tabular}{|c|ccc|}\hline
        Wave & \BF ($10^{-6}$) & $\Re(a_i)$ & $\Im(a_i)$\\ \hline
        S & 11.4 & 1 (fixed) & 0 (fixed)\\
        P & \phantom{0}1.2 & $+3.8$ & \hspace{5pt}$+1.6$\\
        D & 11.4 & $\phantom{+}0.0$ & $-10.4$\\ \hline
    \end{tabular}
    \caption{Branching fraction input with corresponding reverse-engineered couplings for $\Bp \to \rhop \rhoz$.}
    \label{tab:rhoprhoz}
\end{table}

\subsubsection{$\Bz \to \rhop \rhom$}

As before, there should not be a lot of interference from the $a_1(1260)$ resonances to this colour-favoured decay, so the analysis region is kept the same. 
Though the total branching fractions are similar, dilution arising from imperfect flavour-tagging performance taken to be around the 30\% mark for \belletwo means that the yield is set to 30k events. The \CP-violation parameter can be set from the known longitudinal quasi-two-body parameters assuming the solution closest to the SM expectation and uniformity between polarisations. The input branching fractions with the determined couplings and \CP-violation parameters are shown in table~\ref{tab:rhoprhom}.
\begin{table}[!htb]
    \centering
    \begin{tabular}{|c|ccccc|}\hline
        Wave & \BF ($10^{-6}$) & $\Re(a_i)$ & $\Im(a_i)$ & $\Re(\lambda_i)$ & $\Im(\lambda_i)$\\ \hline
        S & 13.7 & 1 (fixed) & 0 (fixed) & $-0.99$ (fixed) & $-0.14$ (fixed)\\
        P & \phantom{0}0.3 & $+1.8$ & $\phantom{+0}0.0$ & " & "\\
        D & 13.7 & $\phantom{+}0.0$ & $-10.4$ & " & "\\ \hline
    \end{tabular}
    \caption{Branching fraction input with corresponding reverse-engineered couplings for $\Bz \to \rhop \rhom$.}
    \label{tab:rhoprhom}
\end{table}

It should also be noted that despite an amplitude analysis being conducted here, it is very unlikely that a single solution for the effective $\phi_2^{+-}$ will emerge. As discussed in ref.~\cite{Dalseno:2018hvf}, a lack of interfering contributions with a sizeable penguin contribution means that the \CP-violation parameter will likely factorise in the isobar sum such that $\Im(\bar A A^*) = \Im(\lambda A A^*) = \Im(\lambda|A|^2)$ in eq.~\ref{eq:tdrate}. As $|A|^2$ must be real-defined, the imaginary part of the aforementioned product evaluates to $\sin 2\phi^{+-}_2$, leaving two solutions remaining for $\phi^{+-}_2$.

\subsubsection{$\Bz \to \rhoz \rhoz$}

According to ref.~\cite{Dalseno:2018hvf}, expanding the analysis space to include $\Bz \to a_1^\pm \pimp$ can ultimately resolve the $\phi_2$ solution degeneracy in $B \to \rho \rho$. Novelty aside, this strategy is prudent as the colour-favoured $\Bz \to a_1^\pm \pimp$ is otherwise either difficult to control systematically or statistically very expensive to remove. As such, the analysis range is defined as the dipion mass being less than $1.1 \gevcc$ as before, or the 3-pion mass being below the production of open charm. 
The combined yield accounting for flavour-tagging performance is estimated in that previous work at 30k events.

\begin{figure}[!htb]
    \centering
    \includegraphics[height=150pt,width=!]{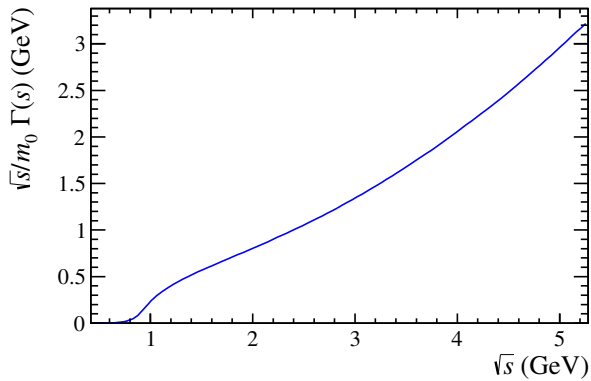}
    \caption{Energy-dependent width of the $a_1^\pm$.}
    \label{fig:a1}
\end{figure}

An additional topology is necessary for $\Bz \to a_1^\pm \pimp$, arising from the S-wave between the products of the $a_1^\pm \to \rho^0 \pipm$ decay,
\begin{equation}
  S \propto L_a(p_{\Bz}, q_{\Bz}) P^{ab}(p_{a_1^\pm}) L_b(p_{\rhoz}, q_{\rhoz}).
\end{equation}
While a relative orbital angular momentum D-wave between the vector and pseudoscalar is possible, this has yet to be definitively seen, so is ignored at this time. Regarding the lineshape of the $a_1^\pm$, potential dispersive effects are neglected, setting $M^2(s)$ to its pole-mass squared. The energy-dependent width of the $a_1^\pm$ is calculated from the integral over its phase space as a function of $s$,
\begin{equation}
    \Gamma_{a_1^\pm}(s) = \frac{1}{2\sqrt{s}} \int \sum_{\lambda=0,\pm 1} |A^\lambda_{a_1^\pm \to (\rho\pi)^\pm_S} (s)|^2 d\Phi_3,
\end{equation}
where $A$ is the transition amplitude of the cascade, itself being comprised of barrier factors, a spin density and lineshape, with a coherent sum taken over the open polarisation indices of the initial state. Its exclusive decay to $(\rho\pi)^\pm$ and isospin symmetry i.e., $\Gamma_{a_1^\pm \to \rhoz \pipm}(s) = \Gamma_{a_1^\pm \to \rho^\pm \piz}(s)$ is also assumed. The numerical form of the $a_1^\pm$ energy-dependent width can be seen in figure~\ref{fig:a1}.

The input branching fractions with the determined couplings and \CP-violation parameters are shown in table~\ref{tab:rhozrhoz}. In this case, only the \lhcb results are used to set the $\Bz \to \rhoz \rhoz$ branching fractions~\cite{Aaij:2015ria}, while the \belle result is used to determine the $\Bz \to a_1^\pm \pimp$ model~\cite{Dalseno:2012hp}.
\begin{table}[!htb]
    \centering
    \begin{tabular}{|c|ccccc|}\hline
        Wave & \BF ($10^{-6}$) & $\Re(a_i)$ & $\Im(a_i)$ & $\Re(\lambda_i)$ & $\Im(\lambda_i)$\\ \hline
        $a_1^+ \pim$ & 8.6 & 1 (fixed) & 0 (fixed) & $-1.04$ (fixed) & $-0.27$ (fixed)\\
        $a_1^- \pip$ & 2.5 & $+0.56$ & $-0.12$ & $-0.80$ (fixed) & $-0.54$ (fixed)\\
        S & 0.4 & $\phantom{+}0.00$ & $+0.01$ & $-0.78$ (fixed) & $+0.25$ (fixed)\\
        P & 0.1 & $+0.08$ & $\phantom{+}0.00$ & " & "\\
        D & 0.4 & $+0.08$ & $+0.09$ & " & "\\ \hline
    \end{tabular}
    \caption{Branching fraction input with corresponding reverse-engineered couplings for $\Bz \to \rhoz \rhoz$ and $\Bz \to a_1^\pm \pimp$.}
    \label{tab:rhozrhoz}
\end{table}

\subsection{Results}

An amplitude fit is performed to the MC sample for each decay in order to determine nominal values for the physics observables of interest against which to compare in the ensuing systematic variations. Three scenarios are defined: ``Current practice'', in which systematic correlations are not considered at all, ``Expected practice'' for which each analysis independently handles their own systematic correlations and finally ``Proposed practice'', where systematic correlations are globally accounted for. Sets of $\rho$ pole parameters are then generated by which to refit the MC samples and calculate covariance between physics observables. Although the \rhop and \rhoz\ parameters are separately determined, they are essentially a manifestation of the same state with different charge, and so for the purposes of evaluating model uncertainties here, are also considered to be fully correlated at the theoretical level. Their parameters taken from ref.~\cite{ParticleDataGroup:2020ssz} are thus distributed with a multivariate normal distribution, recorded in table~\ref{res:rho}.
\begin{table}[!htb]
    \centering
    \begin{tabular}{|c|c|cccc|} \hline
        & Value (GeV$/c^2$) & $m_0$(\rhop) & $\Gamma_0(\rhop)$ & $m_0$(\rhoz) & $\Gamma_0(\rhoz)$\\ \hline
        $m_0$(\rhop) & $0.7665 \pm 0.0011$ & $+1$ & & &\\
        $\Gamma_0(\rhop)/c^2$ & $0.1502	\pm 0.0024$ & $\phantom{+}0$ & $+1$ & &\\
        $m_0$(\rhoz) & $0.7690 \pm 0.0010$ & $+1$  & $\phantom{+}0$ & $+1$ &\\
        $\Gamma_0(\rhoz)/c^2$ & $0.1509	\pm 0.0017$ & $\phantom{+}0$ & $+1$ & $\phantom{+}0$ & $+1$\\ \hline
    \end{tabular}
    \caption{Values and correlation matrix set for the $\rho$ pole parameters in systematic variations.}
    \label{res:rho}
\end{table}

Each MC sample is then refit applying each $\rho$ variation upon which amplitude model covariance matrices are constructed for each scenario. 
Obviously for the Current practice, the amplitude model correlation matrix is set to the identity. In the Expected practice scenario, 100 $\rho$ parameter variations are individually generated for each decay starting with a unique seed, while for the Proposed practice, a single final set of 100 variations is generated to be shared amongst the three analyses. The physics observables included are the branching fractions and where appropriate, the magnitudes of the \CP-violating parameters and their effective weak phases for each polarisation, resulting in a $21 \times 21$ covariance matrix. For completeness, the model uncertainties and their corresponding correlation matrix are given in appendix~\ref{sec:app}.

\begin{figure}[!htb]
    \centering
    \includegraphics[height=139pt,width=!]{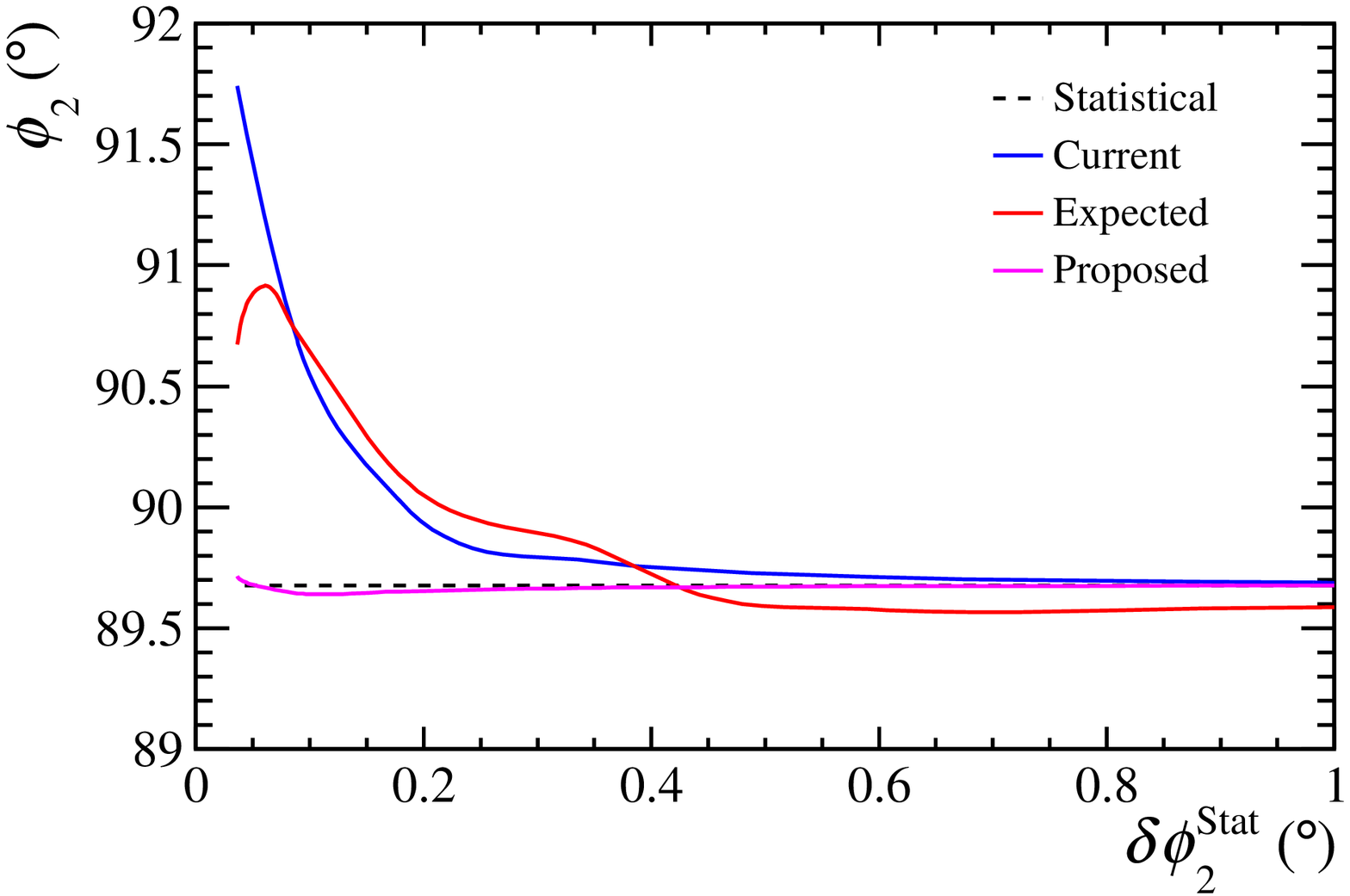}
    \includegraphics[height=139pt,width=!]{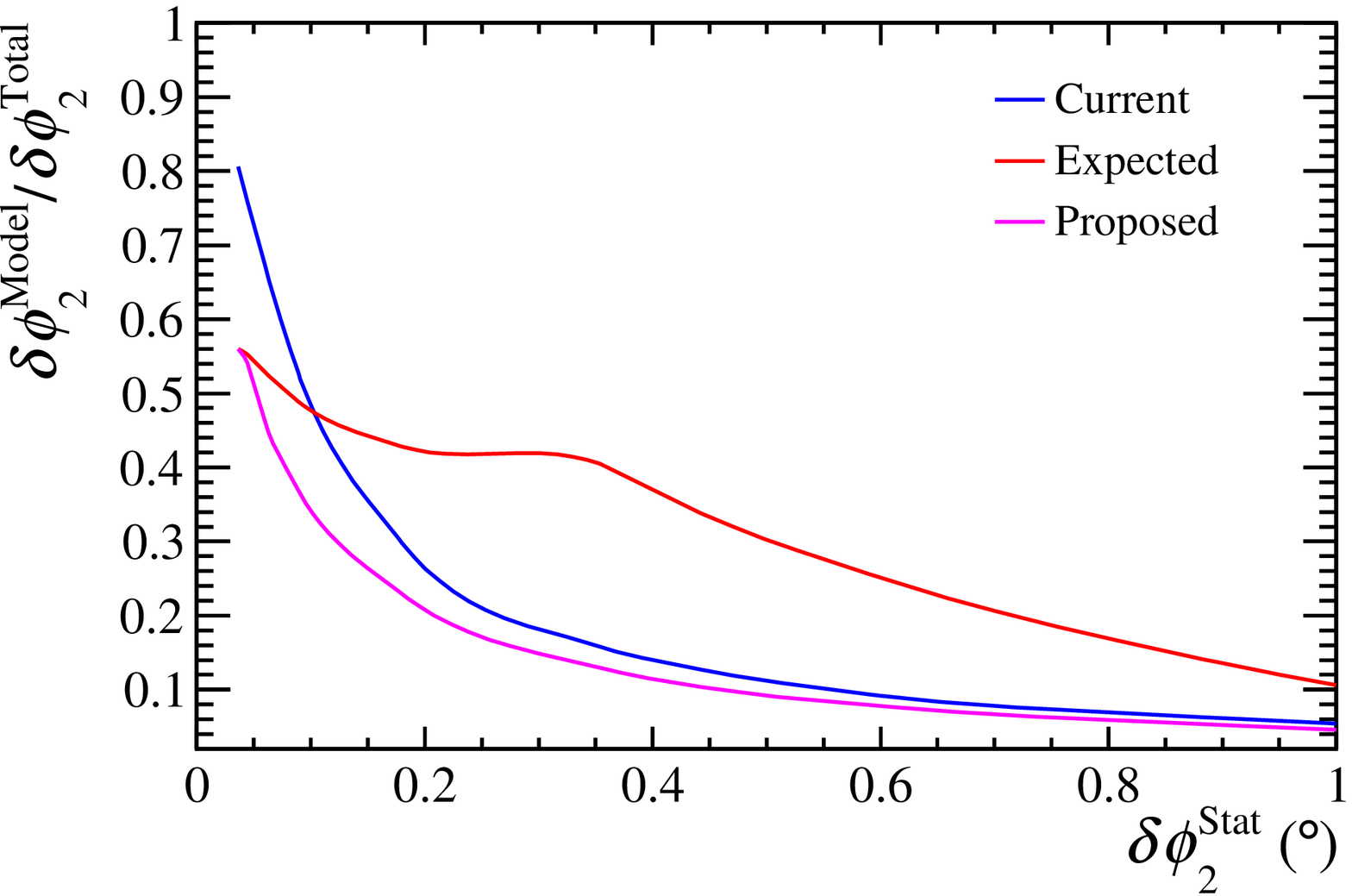}
    \put(-385,115){(a)}
    \put(-175,115){(b)}

    \caption{Performance of the $\phi_2$ constraint under various conditions. (a) shows the $\phi_2$ motion as a function of its statistical uncertainty, while (b) shows the fraction of the amplitude model uncertainty.}
    \label{fig:rhocorr1}
\end{figure}

The $\phi_2$ constraint summing over all polarisations is then conducted in each scenario. The model uncertainty is not taken to scale with statistics considering the large yields involved because the variations of the $\rho$ parameters change the underlying interference pattern in the phase space in a predetermined albeit unknown way, which is why this systematic is irreducible without appreciable improvements to the pole properties themselves. As such, I repeat the $\phi_2$ constraint in the context of a changing statistical uncertainty which is achieved by scaling the statistical covariance matrix obtained from the nominal fits, with the results shown in figure~\ref{fig:rhocorr1}.

The raw drift of $\phi_2$ can be seen in figure~\ref{fig:rhocorr1}a. For reference, a statistical-only constraint ignoring the amplitude model covariance matrix is performed for which $\phi_2$ is perfectly flat as a function of its own overall statistical uncertainty as expected. For these values of statistical uncertainty, the trends of each analysis scenario are then determined. When accounting for the model uncertainty, but not any corresponding correlations, the constraint has no problem at the statistical precision of $1^\circ$, but rapidly deteriorates to become the worst performer once the model uncertainty begins to dominate. 
Surprisingly, the Expected practice scenario in which each analysis considers their own model correlations already has a visible bias at $1^\circ$ uncertainty, trending in the same direction as Current practice but without any intuitively discernible motion. Finally, the Proposed practice curve tightly follows the statistical-only curve, indicating that its covariance matrix by and large captures the correct structure of the model uncertainty. Perhaps the slight departure from the flat curve at low values indicates the point at which sensitivity to non-linear correlations begins to play a role.

Figure~\ref{fig:rhocorr1}b indicates the size of the model uncertainty as a percentage of the total uncertainty in $\phi_2$, which is obtained by quadratically subtracting the statistical uncertainty at that point, thereby assuming that all other systematics behave like the statistical error. Current practice performs well when the statistical uncertainty dominates, but becomes overtly large at the other end of the spectrum. Conversely, the model uncertainty fraction for Expected practice is noticeably larger even where the statistical uncertainty is supposed to be dominant, while its performance even improves to the point of the Proposed practice though with substantial bias as already seen.

\begin{figure}[!htb]
    \centering
    \includegraphics[height=139pt,width=!]{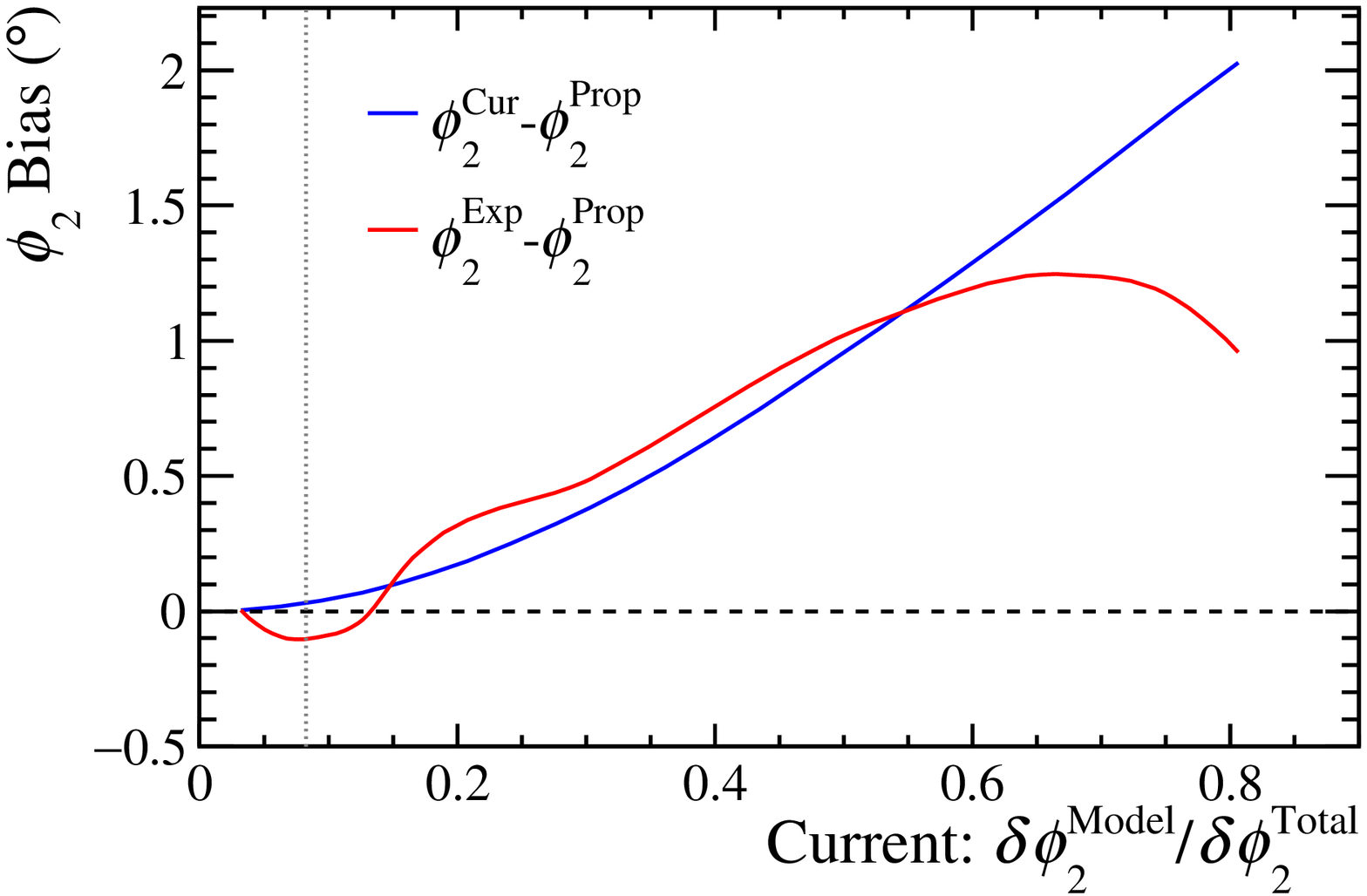}
    \includegraphics[height=139pt,width=!]{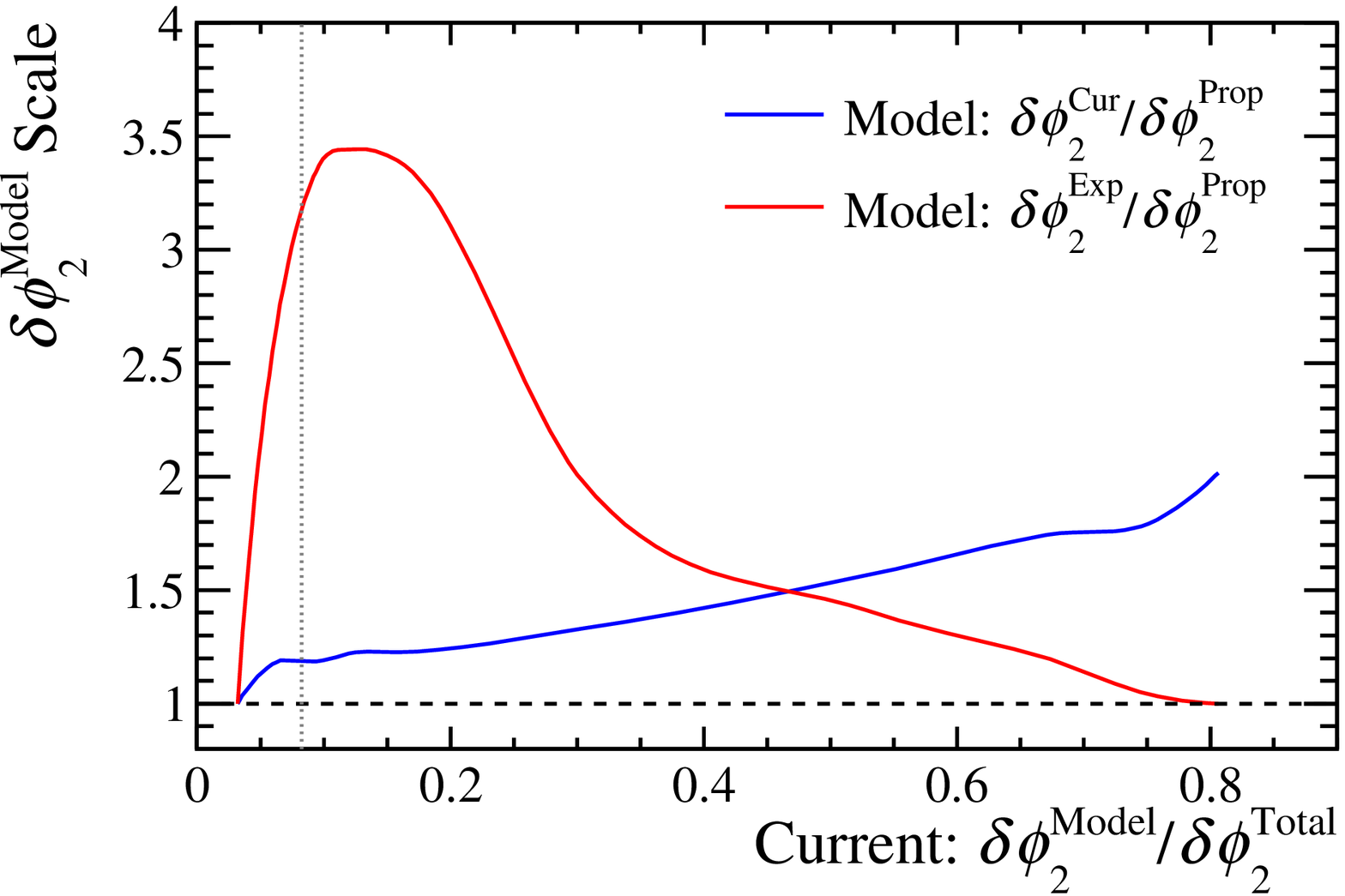}
    \put(-245,30){(a)}
    \put(-30,35){(b)}

    \caption{Performance of the $\phi_2$ constraint under various conditions. Under Current practice handling of the model uncertainty fraction, the bias in $\phi_2$ relative to proposed practice is shown in (a), while the degradation of the model uncertainty size is shown in (b). The vertical dotted line corresponds to the \belletwo\ projection point for their full data sample.}
    \label{fig:rhocorr2}
\end{figure}

\begin{figure}
    \centering
    \includegraphics[height=150pt,width=!]{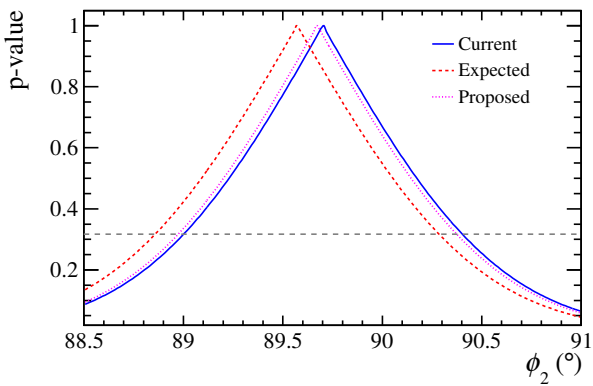}
    \caption{$p$-value scan of $\phi_2$ where the horizontal dashed line shows the $1\sigma$ bound. The blue, red and magenta curves show the Current, Expected and Proposed practice scenarios, respectively at the point where the statistical covariance matrix is aligned to give the total uncertainty expected by the end of \belletwo.}
    \label{fig:rhorho}
\end{figure}

The plots of figure~\ref{fig:rhocorr2} are variations of those already shown, though now as a function of the model uncertainty fraction in the Current practice scenario and relative to Proposed practice, which allows the difference in performance to be understood at the \belletwo projection of $0.7^\circ$ total uncertainty with $50 \ab^{-1}$~\cite{Kou:2018nap}. This is achieved by tuning the statistical covariance matrix scale factor, the result of which is shown in figure~\ref{fig:rhorho}. Back in figure~\ref{fig:rhocorr2}a, the bias in $\phi_2$ with the full \belletwo dataset would be only $+0.03^\circ$ with Current practice, but $-0.1^\circ$ with Expected practice for the studied model. 
Figure~\ref{fig:rhocorr2}b indicates how much worse the model uncertainty scales as a function of its fraction, showing that at \belletwo uncertainties, Current practice is worse-off than Proposed practice only by factor of 1.2, while Expected practice has a model uncertainty 3 times worse.

For the given amplitude models studied, it would appear that Proposed practice is objectively superior. Current practice may initially perform well, but is expected to become problematic, while Expected practice just looks dangerous from the onset, which is perhaps an unexpected outcome. A natural question arising at this point is whether this whole issue can be sidestepped by releasing the $\rho$ pole properties in the fit. However, the meaning of the $\phi_2$ constraint would be unclear with multiple versions of these parameters present. In principle, it would be possible to release the $\rho$ pole parameters in a simultaneous fit to the three $B \to \rho \rho$ modes, trading relative analysis simplicity for easier sytematics handling, though understandably this alternative is not very practical in terms of coordination and the short-term contractual nature of the field. 
In any case, there is more to each amplitude model than just the $B \to \rho\rho$ contributions as mentioned at the beginning of this section, so the full correlated model uncertainty accounting for other common lineshapes and additional contributions should also be studied. Minimal cooperation and overlap between analyses through the sharing of systematic variations and their signed fit residuals would seem to be the most sensible strategy moving forwards.

%% file: rhopi.tex
\section{Amplitude model correlations between systems}
\label{sec:rhopi}

Now that systematic bias in $\phi_2$ and precision being left on the table is found to occur when not considering amplitude model correlations properly, or indeed even at all, attention turns to the other system also involving the $\rho$ lineshape in an amplitude analysis, $\Bz \to (\rho \pi)^0$. While this involves a single analysis in which the complete set of systematic uncertainties are already considered as standard, the question remains whether the global treatment of amplitude model correlations in $B \to \rho\rho$ is sufficient, or whether its scope should also be expanded to include $\Bz \to (\rho \pi)^0$. Recalling the discussion of section~\ref{sec:nbb}, this is analogous to the realisation that all branching fractions of the $B \to \pi \pi$ and $\rho \rho$ systems are also systematically correlated through $N_{B\Bbar}$.

\subsection{Amplitude model}

The overlap of the three $\Bz \to (\rho \pi)^0$ charge combinations in the $\Bz \to \pip \pim \piz$ phase space allows the tree ($T$) and penguin ($P$) processes involved to be distinguished, leading to the direct measurement of $\phi_2$ in a single analysis~\cite{Snyder:1993mx}. Additional constraints from isospin pentagonal relations involving the charged $B$ modes can also help to improve the constraint~\cite{Lipkin:1991st}, though this extension will not be addressed further here, suffice it to say that a global approach to amplitude model correlations in the wider $B \to \rho \pi$ system will likely be needed as well in light of what has been seen so far. The single analysis applies the isospin symmetry argument to the penguin amplitudes rather than the tree as is the case in other systems, lessening the impact of theoretical uncertainties~\cite{Gronau:2005pq}. 
The decomposition of the complex amplitude couplings by charge and $B$ flavour is given in eq.~\ref{eq:rhopi},
\begin{alignat}{6}
    &A^{+-} &&= T^{+-}e^{-i\phi_2} &&+ P^{+-}, \hspace{20pt} &&\bar A^{-+} &&= T^{+-}e^{+i\phi_2} &&+ P^{+-}, \nonumber \\
    &A^{-+} &&= T^{-+}e^{-i\phi_2} &&+ P^{-+}, \hspace{20pt} &&\bar A^{+-} &&= T^{-+}e^{+i\phi_2} &&+ P^{-+}, \nonumber \\
    &A^{00} &&= T^{00}e^{-i\phi_2} &&+ \frac{1}{2}(P^{+-} + P^{-+}), \hspace{20pt} &&\bar A^{00} &&= T^{00}e^{+i\phi_2} &&+ \frac{1}{2}(P^{+-} + P^{-+}),
    \label{eq:rhopi}
\end{alignat}
where the superscripts represent the $\rho$ followed by the pion charge. It can be seen that sensitivity to the penguin amplitudes amongst the trees arises as a result of the isospin argument removing the independence of the colour-suppressed $\Bz \to \rhoz \piz$ penguin.

However, by construction the tree and penguin couplings are highly correlated, so a parameterisation expanding the product of isobar sums for each of the three terms in eq.~\ref{eq:tdrate} was proposed~\cite{Quinn:2000by}. For each resulting amplitude-squared-level form factor composed solely of strong dynamics, the expression for the corresponding coupling combination is then substituted with an independent free parameter. The parameter space thus increases dramatically in return for statistical stability of the fit, after which a $\chi^2$ minimisation mapping the original tree and penguin amplitudes to the 27 bilinear coefficients can be executed to recover $\phi_2$. At this time, it is unclear if this method will persist going forwards due to the unphysical assumption that the higher $\rho$ resonances are imposed to share the same $P/T$ ratios and the related impracticalities in adding further structures to the amplitude model.

For the purposes of this study, since the tree and penguin amplitudes are known from MC generation, the fit will be performed with that paradigm for simplicity as $\phi_2$ will be a free parameter. The input amplitudes themselves are still obtained from a $\chi^2$ fit to the Belle bilinear coefficients, taking the solution consistent with SM expectations~\cite{Kusaka:2007dv,Kusaka:2007mj}. The parameters are given in table~\ref{tab:rhopi}.
\begin{table}[!htb]
    \centering
    \begin{tabular}{|c|c|} \hline
        Parameter & Value \\ \hline
        $\phi_2$ & $81.4^\circ$ \\
        $\Re(T^{+-})$ & $+0.80$ (fixed)\\
        $\Im(T^{+-})$ & $\phantom{+}0\phantom{.00}$ (fixed)\\
        $\Re(T^{-+})$ & $+0.56$\\
        $\Im(T^{-+})$ & $+0.11$\\
        $\Re(T^{00})$ & $+0.07$\\
        $\Im(T^{00})$ & $+0.37$\\
        $\Re(P^{+-})$ & $+0.20$\\
        $\Im(P^{+-})$ & $+0.10$\\
        $\Re(P^{-+})$ & $-0.30$\\
        $\Im(P^{-+})$ & $-0.01$\\ \hline
    \end{tabular}
    \caption{MC input for $\Bz \to (\rho \pi)^0$.}
    \label{tab:rhopi}
\end{table}

\subsection{Results}

The $\Bz \to (\rho \pi)^0$ effective yield accounting for flavour-tagging dilution is set at 30k events for the $50 \ab^{-1}$ expected at \belletwo, with two scenarios considered for systematic variations. The first is that each system is treated independently, with $B \to \rho\rho$ already adopting the globally treated correlations proposed in the previous section as standard, while in the second, the systematic variations are common to both systems. The model uncertainty on $\phi_2$ and how it correlates with the $B \to \rho \rho$ parameters are again given in appendix~\ref{sec:app}. A $\phi_2$ scan is performed over the lesser known quantities, which again includes the strength of the $B \to \rho \rho$ model uncertainty in that system and the total uncertainty in $\Bz \to (\rho \pi)^0$ relative to $B \to \rho \rho$.

\begin{figure}[!htb]
    \centering
    \includegraphics[height=139pt,width=!]{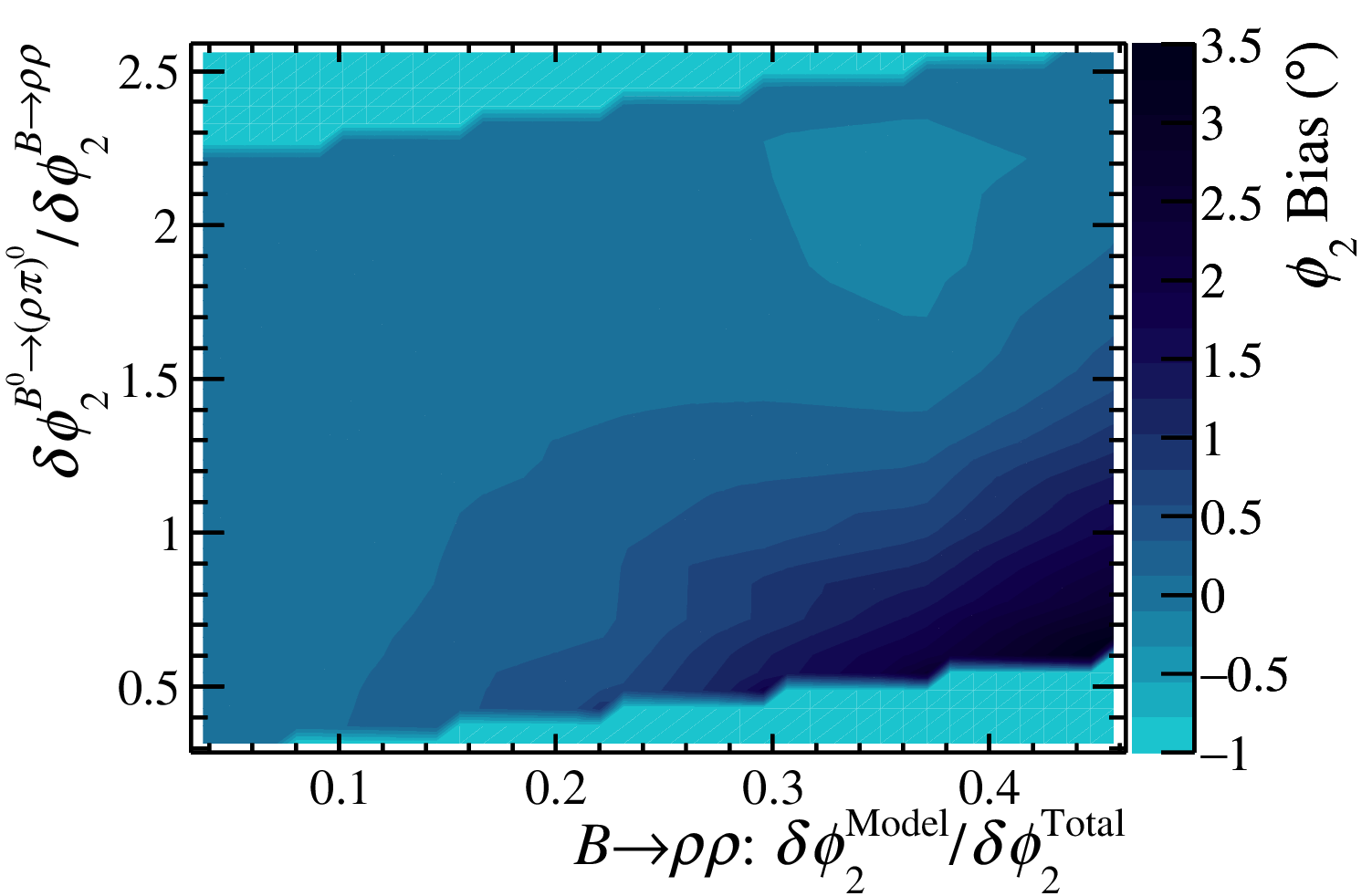}
    \includegraphics[height=139pt,width=!]{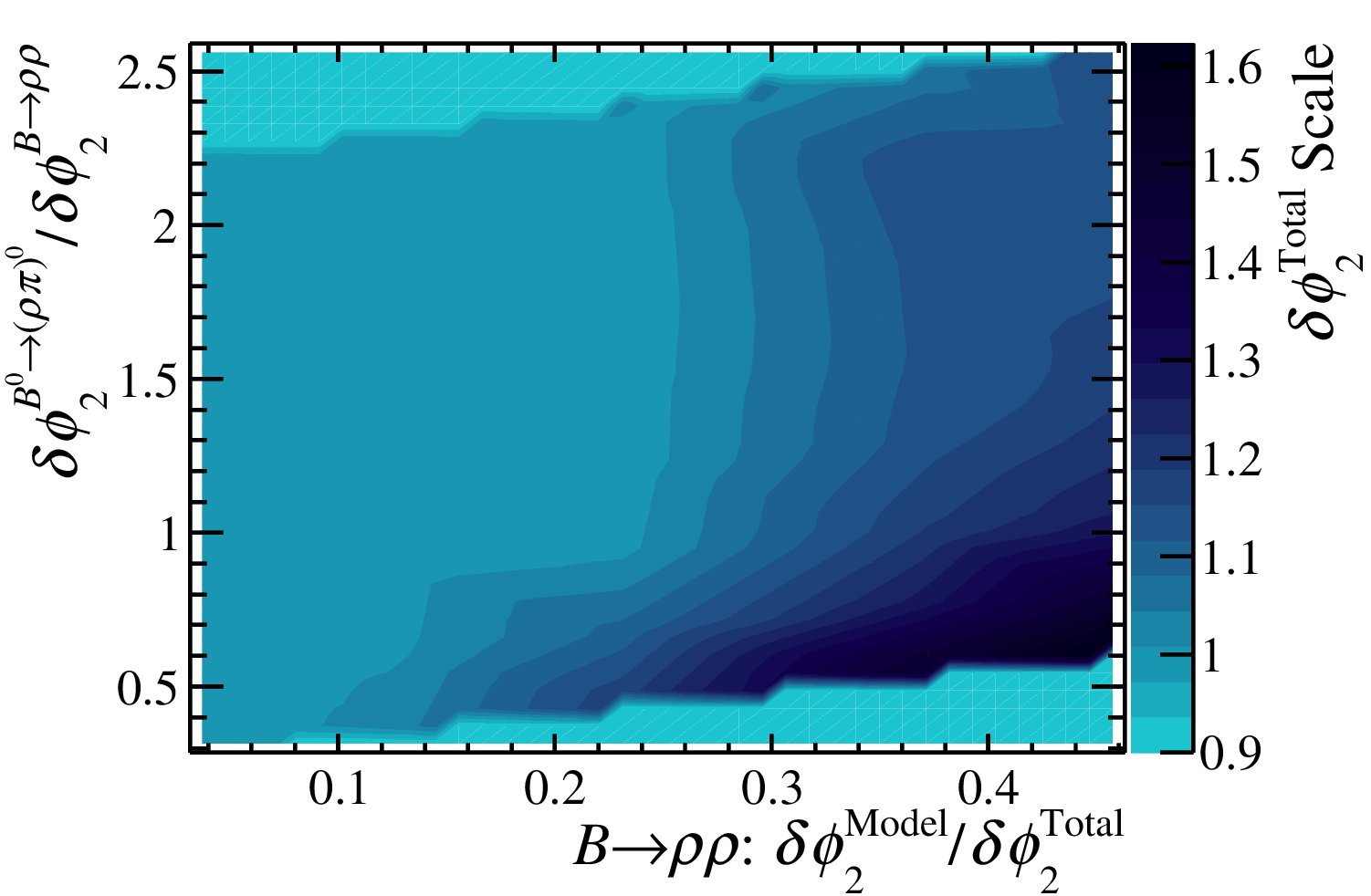}
    \put(-390,120){(a)}
    \put(-174,120){(b)}

    \caption{Performance of the $\phi_2$ constraint under various conditions. The bias in $\phi_2$ when treating amplitude model correlations in $B \to \rho \rho$ and $\Bz \to (\rho \pi)^0$ separately as opposed to globally is shown in (a), while the degradation of the total uncertainty in $\phi_2$ is shown in (b). The jagged edges indicate the limits of the scan, outside of which the contents can be ignored.}
    \label{fig:rhopicorr}
\end{figure}

Figure~\ref{fig:rhopicorr}a shows the scope of the bias in $\phi_2$, while Figure~\ref{fig:rhopicorr}b demonstrates the loss in statistical power when treating each system independently. From these plots, it is clear that a coordinated approach to these analyses would be required if $\Bz \to (\rho \pi)^0$ dominates the precision when at the same time a sizeable model uncertainty is present in the $B \to \rho \rho$ system.

\subsection{A word on amplitude model correlations between experiments}

So far, no mention has been made on the role of \lhcb, which at this time is expected to provide only partial input to the $\phi_2$ constraint, primarily from the $\Bz \to \rhoz \rhoz$ decay. Depending on the statistical power of this analysis, which will be considerable, it has already been shown that each analysis and therefore experiment by extension, when left to their own devices concerning the handling of amplitude model correlations, is most detrimental to the average.

The logical heresy is to combine \belletwo and \lhcb data in a single analysis and jointly handle the systematics. In lieu of this ideal scenario, which is understandably fraught with political difficulties, an unbiased outcome can still be achieved without the sharing of data sets through rigorous bookkeeping. These experiments can communicate with each other to  define the $\rho$ pole parameters amongst others and provide a standard set of variations for all analyses to use. In return, each analysis can report the signed fit residual obtained for each systematic variation on all physics observables measured so that systematic covariance matrices can be properly constructed.

%% file: conclusion.tex
\section{Conclusion}
\label{sec:conclusion}

One of the many challenges in $\phi_2$-sensitive studies is that similarity of the final states and overlapping analysis techniques inevitably leads to significant systematic correlations amongst the physics observables ultimately constraining $\phi_2$. A coordinated approach leads to an appreciably overall improved precision, while at the same time eliminating bias, which in the case of $B \to \rho\rho$ is shown to be at the level of $1^\circ$ in and of itself. Additional care may also need to be taken if $\Bz \to (\rho \pi)^0$ turns out to dominate the average whilst at the same time, the amplitude model uncertainty in $B \to \rho\rho$ is significant. Physics parameter correlations in data models can cross experimental lines for which minimal cooperation through the sharing of signed fit residuals on commonly defined systematic variations offers a good compromise to combined data analyses. Hopefully, this work inspires other investigations into the role of systematic correlations beyond single measurements in the combinations of other \CP-violating weak phases such as $\phi_1$ ($\beta$), $\phi_3$ ($\gamma$) and $\phi_s$.

%% file: covariance.tex
\section{Model uncertainties and correlation matrix}
\label{sec:app}

For the considered $B \to \rho \rho$ and $\Bz \to (\rho \pi)^0$ amplitude models, the systematic uncertainties on the extracted physics parameters constraining $\phi_2$ that globally account for correlations arising from the $\rho$ pole parameters are given in table~\ref{tab:syst}. For comparison, how these uncertainties scale when ignoring such correlations within and between these systems is also shown. The associated correlation matrix in the global case is spread over tables~\ref{tab:corr1}-\ref{tab:corr3}, while the uncorrelated scenarios can be trivially inferred from these.

\begin{table}[!htb]
    \centering
    \begin{tabular}{|c|c|c|} \hline
        Parameter & Model Uncertainty Proposed & Uncorrelated Scale\\ \hline
        $\mathcal{B}^{00}$ [S] & $3.1 \%$ & 1.10 \\
        $|\lambda_{\CP}^{00}|$ [S] & 0.021 & 0.94\\
        $\phi_2^{00}$ [S] & $0.51^\circ$ & 0.99\\
        $\mathcal{B}^{00}$ [P] & $0.090 \%$ & 1.13\\
        $|\lambda_{\CP}^{00}|$ [P] & 0.0010 & 1.14\\
        $\phi_2^{00}$ [P] & $0.16^\circ$ & 1.12\\
        $\mathcal{B}^{00}$ [D] & $1.39 \%$ & 1.09\\
        $|\lambda_{\CP}^{00}|$ [D] & 0.039 & 1.00\\
        $\phi_2^{00}$ [D] & $0.53^\circ$ & 1.00\\
        $\mathcal{B}^{+-}$ [S] & $0.26 \%$ & 1.05\\
        $|\lambda_{\CP}^{+-}|$ [S] & 0.00045 & 1.11\\
        $\phi_2^{+-}$ [S] & $0.019^\circ$ & 1.07\\
        $\mathcal{B}^{+-}$ [P] & $0.45 \%$ & 1.07\\
        $|\lambda_{\CP}^{+-}|$ [P] & 0.0013 & 1.05\\
        $\phi_2^{+-}$ [P] & $0.022^\circ$ & 1.12\\
        $\mathcal{B}^{+-}$ [D] & $1.7\%$ & 1.07\\
        $|\lambda_{\CP}^{+-}|$ [D] & 0.0018 & 1.07\\
        $\phi_2^{+-}$ [D] & $0.015^\circ$ & 1.11\\
        $\mathcal{B}^{+0}$ [S] & $0.25 \%$ & 0.85\\
        $\mathcal{B}^{+0}$ [P] & $0.11 \%$ & 0.81\\
        $\mathcal{B}^{+0}$ [D] & $1.55 \%$ & 0.78\\
        $\phi_2$ [$\Bz \to (\rho \pi)^0$] & $0.081^\circ$ & 1.03\\ \hline
    \end{tabular}
    \caption{Model uncertainties under the Proposed practice scheme for each determined parameter entering the $\phi_2$ constraint. The final column shows how the uncertainties scale when correlations within and between systems are ignored. Apart from the final row entry, square brackets indicate the partial wave in the $B \to \rho \rho$ sector.}
    \label{tab:syst}
\end{table}

\begin{sidewaystable}[!htb]
    \centering
    \begin{tabular}{|c|ccccccccc|} \hline
        & $\mathcal{B}^{00}$ [S] & $|\lambda_{\CP}^{00}|$ [S] & $\phi_2^{00}$ [S] &
        $\mathcal{B}^{00}$ [P] & $|\lambda_{\CP}^{00}|$ [P] & $\phi_2^{00}$ [P] &
        $\mathcal{B}^{00}$ [D] & $|\lambda_{\CP}^{00}|$ [D] & $\phi_2^{00}$ [D]\\ \hline

        $\mathcal{B}^{00}$ [S] & $+1.00$ &&&&&&&&\\
        $|\lambda_{\CP}^{00}|$ [S] & $-0.88$ & $+1.00$ &&&&&&&\\
        $\phi_2^{00}$ [S] & $+0.85$ & $-0.99$ & $+1.00$ &&&&&&\\

        $\mathcal{B}^{00}$ [P] & $-0.78$ & $+0.45$ & $-0.36$ & $+1.00$ &&&&&\\
        $|\lambda_{\CP}^{00}|$ [P] & $+0.61$ & $-0.22$ & $+0.12$ & $-0.97$ & $+1.00$ &&&&\\
        $\phi_2^{00}$ [P] & $-0.49$ & $+0.09$ & $+0.02$ & $+0.92$ & $-0.99$ & $+1.00$ &&&\\

        $\mathcal{B}^{00}$ [D] & $+1.00$ & $-0.90$ & $+0.87$ & $-0.76$ & $+0.58$ & $-0.45$ & $+1.00$ &&\\
        $|\lambda_{\CP}^{00}|$ [D] & $-0.96$ & $+0.97$ & $-0.94$ & $+0.66$ & $-0.45$ & $+0.32$ & $-0.97$ & $+1.00$ &\\
        $\phi_2^{00}$ [D] & $-0.97$ & $+0.89$ & $-0.83$ & $+0.80$ & $-0.64$ & $+0.53$ & $-0.97$ & $+0.97$ & $+1.00$\\

        $\mathcal{B}^{+-}$ [S] & $+0.87$ & $-0.54$ & $+0.48$ & $-0.98$ & $+0.92$ & $-0.85$ & $+0.85$ & $-0.74$ & $-0.85$\\
        $|\lambda_{\CP}^{+-}|$ [S] & $+0.65$ & $-0.33$ & $+0.23$ & $-0.95$ & $+0.95$ & $-0.93$ & $+0.63$ & $-0.54$ & $-0.70$\\
        $\phi_2^{+-}$ [S] & $-0.80$ & $+0.45$ & $-0.38$ & $+0.96$ & $-0.93$ & $+0.87$ & $-0.77$ & $+0.66$ & $+0.79$\\

        $\mathcal{B}^{+-}$ [P] & $-0.78$ & $+0.43$ & $-0.35$ & $+1.00$ & $-0.97$ & $+0.93$ & $-0.76$ & $+0.64$ & $+0.79$\\
        $|\lambda_{\CP}^{+-}|$ [P] & $-0.85$ & $+0.54$ & $-0.47$ & $+0.99$ & $-0.93$ & $+0.87$ & $-0.83$ & $+0.74$ & $+0.86$\\
        $\phi_2^{+-}$ [P] & $+0.54$ & $-0.15$ & $+0.04$ & $-0.94$ & $+0.99$ & $-0.99$ & $+0.51$ & $-0.39$ & $-0.58$\\

        $\mathcal{B}^{+-}$ [D] & $-0.80$ & $+0.46$ & $-0.37$ & $+1.00$ & $-0.96$ & $+0.91$ & $-0.77$ & $+0.67$ & $+0.81$\\
        $|\lambda_{\CP}^{+-}|$ [D] & $+0.80$ & $-0.46$ & $+0.38$ & $-0.99$ & $+0.96$ & $-0.91$ & $+0.78$ & $-0.66$ & $-0.80$\\
        $\phi_2^{+-}$ [D] & $+0.63$ & $-0.30$ & $+0.20$ & $-0.95$ & $+0.95$ & $-0.94$ & $+0.60$ & $-0.51$ & $-0.68$\\

        $\mathcal{B}^{+0}$ [S] & $+0.86$ & $-0.53$ & $+0.47$ & $-0.98$ & $+0.92$ & $-0.86$ & $+0.84$ & $-0.73$ & $-0.84$\\
        $\mathcal{B}^{+0}$ [P] & $-0.77$ & $+0.43$ & $-0.34$ & $+1.00$ & $-0.97$ & $+0.93$ & $-0.75$ & $+0.64$ & $+0.79$\\
        $\mathcal{B}^{+0}$ [D] & $-0.81$ & $+0.48$ & $-0.40$ & $+1.00$ & $-0.96$ & $+0.90$ & $-0.79$ & $+0.69$ & $+0.82$\\

        $\phi_2$ [$\Bz \to (\rho \pi)^0$] & $-0.98$ & $+0.76$ & $-0.73$ & $+0.87$ & $-0.74$ & $+0.63$ & $-0.97$ & $+0.90$ & $+0.94$\\ \hline
    \end{tabular}
    \caption{Correlation matrix (1) under the Proposed practice scheme for each determined parameter entering the $\phi_2$ constraint. Apart from the final row entry, square brackets indicate the partial wave in the $B \to \rho \rho$ sector.}
    \label{tab:corr1}
\end{sidewaystable}

\begin{sidewaystable}[!htb]
    \centering
    \begin{tabular}{|c|ccccccccc|} \hline
        & $\mathcal{B}^{+-}$ [S] & $|\lambda_{\CP}^{+-}|$ [S] & $\phi_2^{+-}$ [S] &
        $\mathcal{B}^{+-}$ [P] & $|\lambda_{\CP}^{+-}|$ [P] & $\phi_2^{+-}$ [P] &
        $\mathcal{B}^{+-}$ [D] & $|\lambda_{\CP}^{+-}|$ [D] & $\phi_2^{+-}$ [D]\\ \hline

        $\mathcal{B}^{+-}$ [S] & $+1.00$ &&&&&&&&\\
        $|\lambda_{\CP}^{+-}|$ [S] & $+0.89$ & $1.00$ &&&&&&&\\
        $\phi_2^{+-}$ [S] & $-0.96$ & $-0.85$ & $+1.00$ &&&&&&\\

        $\mathcal{B}^{+-}$ [P] & $-0.98$ & $-0.94$ & $+0.97$ & $+1.00$ &&&&&\\
        $|\lambda_{\CP}^{+-}|$ [P] & $-0.99$ & $-0.92$ & $+0.96$ & $+0.99$ & $+1.00$ &&&&\\
        $\phi_2^{+-}$ [P] & $+0.87$ & $+0.96$ & $-0.88$ & $-0.94$ & $-0.89$ & $+1.00$ &&&\\

        $\mathcal{B}^{+-}$ [D] & $-0.99$ & $-0.94$ & $+0.97$ & $+1.00$ & $+0.99$ & $-0.93$ & $+1.00$ &&\\
        $|\lambda_{\CP}^{+-}|$ [D] & $+0.99$ & $+0.91$ & $-0.99$ & $-0.99$ & $-0.99$ & $+0.92$ & $-1.00$ & $+1.00$ &\\
        $\phi_2^{+-}$ [D] & $+0.89$ & $+1.00$ & $-0.85$ & $-0.94$ & $-0.92$ & $+0.97$ & $-0.94$ & $+0.91$ & $+1.00$\\

        $\mathcal{B}^{+0}$ [S] & $+1.00$ & $+0.90$ & $-0.97$ & $-0.99$ & $-1.00$ & $+0.88$ & $-0.99$ & $+0.99$ & $+0.89$\\
        $\mathcal{B}^{+0}$ [P] & $-0.98$ & $-0.94$ & $+0.97$ & $+1.00$ & $+0.99$ & $-0.95$ & $+1.00$ & $-0.99$ & $-0.94$\\
        $\mathcal{B}^{+0}$ [D] & $-0.99$ & $-0.93$ & $+0.97$ & $+1.00$ & $+1.00$ & $-0.92$ & $+1.00$ & $-1.00$ & $-0.93$\\

        $\phi_2$ [$\Bz \to (\rho \pi)^0$] & $-0.94$ & $-0.75$ & $+0.88$ & $+0.87$ & $+0.93$ & $-0.67$ & $+0.89$ & $-0.89$ & $-0.73$\\ \hline
    \end{tabular}
    \caption{Correlation matrix (2) under the Proposed practice scheme for each determined parameter entering the $\phi_2$ constraint. Apart from the final row entry, square brackets indicate the partial wave in the $B \to \rho \rho$ sector.}
    \label{tab:corr2}
\end{sidewaystable}

\begin{table}[!htb]
    \centering
    \begin{tabular}{|c|cccc|} \hline
        & $\mathcal{B}^{+0}$ [S] & $\mathcal{B}^{+0}$ [P] & $\mathcal{B}^{+0}$ [D] &
        $\phi_2$ [$\Bz \to (\rho \pi)^0$]\\ \hline

        $\mathcal{B}^{+0}$ [S] & $+1.00$ &&&\\
        $\mathcal{B}^{+0}$ [P] & $-0.98$ & $+1.00$ &&\\
        $\mathcal{B}^{+0}$ [D] &$-0.99$ & $+1.00$ & $+1.00$ &\\

        $\phi_2$ [$\Bz \to (\rho \pi)^0$] & $-0.94$ & $+0.87$ & $+0.90$ & $+1.00$\\ \hline
    \end{tabular}
    \caption{Correlation matrix (3) under the Proposed practice scheme for each determined parameter entering the $\phi_2$ constraint. Apart from the final row entry, square brackets indicate the partial wave in the $B \to \rho \rho$ sector.}
    \label{tab:corr3}
\end{table}